\documentclass[%
reprint,
superscriptaddress,
amsmath,amssymb,
aps,
pra,
]{revtex4-2}

\usepackage{graphicx}
\usepackage{dcolumn}
\usepackage{bm}
\usepackage[bookmarks=true, hidelinks, bookmarksopen, bookmarksopenlevel=1]{hyperref}
\usepackage[separate-uncertainty=true]{siunitx} 
\DeclareSIUnit{\au}{a.u.}
\usepackage[font=normal]{subcaption}
\usepackage{xcolor}
\usepackage{ragged2e}
\usepackage{ulem}
\usepackage{physics}
\usepackage{csquotes} 
\usepackage{url}

\renewcommand{\emph}[1]{\textit{#1}}

\begin{document}

    \preprint{APS/123-QED}

    \title{Entanglement and photoelectron holography in dissociative photoionization: molecular quantum eraser}
    
    \author{Sebastian Hell}
    \email{sebastian.hell@uni-jena.de}
    \affiliation{Institute for Optics and Quantum Electronics, Friedrich Schiller University, 07743 Jena, Germany}
    
    \author{Paul Winter}
    \affiliation{Institute of Theoretical Physics, Leibniz University Hannover, 30167 Hannover, Germany}

    \author{Martin Gärttner}
    \affiliation{Institute of Condensed Matter Theory and Optics, Friedrich Schiller University, 07743 Jena, Germany}
    
    \author{Julian Späthe}
    \affiliation{Institute for Optics and Quantum Electronics, Friedrich Schiller University, 07743 Jena, Germany}
    
    \author{Saurabh Mhatre}
    \affiliation{Institute for Optics and Quantum Electronics, Friedrich Schiller University, 07743 Jena, Germany}
    
    \author{Dejan B. Milo\v{s}evi\'{c}}
    \affiliation{University of Sarajevo, Faculty of Science, 71000 Sarajevo, Bosnia and Herzegovina}
    \affiliation{Academy of Sciences and Arts of Bosnia and Herzegovina, 71000 Sarajevo, Bosnia and Herzegovina}
    
    \author{Gerhard G. Paulus}
    \affiliation{Institute for Optics and Quantum Electronics, Friedrich Schiller University, 07743 Jena, Germany}
    \affiliation{Helmholtz Institute Jena, 07743 Jena, Germany}
    
    \author{Manfred Lein}
    \affiliation{Institute of Theoretical Physics, Leibniz University Hannover, 30167 Hannover, Germany}
    
    \author{Matthias K\"ubel}
    \affiliation{Institute for Optics and Quantum Electronics, Friedrich Schiller University, 07743 Jena, Germany}
    \affiliation{Helmholtz Institute Jena, 07743 Jena, Germany}

    \date{\today}
    
    \begin{abstract}

    In a double-slit experiment with a bipartite system, the visibility of interference fringes depends on the availability of which-way information.
    Here, we report a Bell-like state of photoelectron and residual ion in the multiphoton dissociative ionization of the D$_2$ molecule. Evidence for entanglement is provided by the correlated emission directions of photoelectron and ion, which is observed using a COLTRIMS reaction microscope. In the presence of this correlation, the holographic interference fringes contained in the photoelectron momentum distributions are suppressed, indicating the existence of which-way information. The interference pattern persists when a single ionic state is populated and thus no which-way information is available. The experimental observations and conclusions are supported by the numerical solution of the electronic-nuclear time-dependent Schrödinger equation.   
    Our work demonstrates that coincidence spectroscopy of ions and electrons is a powerful method for studying fundamental concepts of quantum information science within the context of ultrafast light-matter interactions. 

    \end{abstract}
    
    \maketitle
    
  
    The quantum eraser experiment \cite{Scully1982quantum,Scully1991,Kwiat1992Observation} impressively demonstrates that the results of measurements depend on what information is available about a quantum system. The coherence of a particle is verified by letting it interfere with itself in a double-slit configuration. When the particle becomes entangled with an auxiliary \textquote{marker} particle that carries information about the interfering pathways, the interference pattern vanishes, which suggests the loss of coherence. However, the total system of the two entangled particles remains in a pure and thus coherent state. The interference fringes can be recovered through a coincidence measurement of both particles in a basis that masks the which-way information, and suitable postselection of the results. Such quantum eraser experiments have been realized in a variety of systems, ranging from photons \cite{Kwiat1992Observation,Walborn2002} and phonons \cite{Bienfait2020} to electrons \cite{Weisz2014} and atoms \cite{Duerr1998}. 

    In the context of ultrafast photoionization, entanglement phenomena such as coherence, correlation, control thereof, and measures of entanglement, have been studied rather extensively using theoretical methods \cite{fedorov2004packet, spanner2007entanglement, rohringer2009multichannel,  majorosi2017quantum, nishi2019entanglement, ruberti2019onset, vrakking2021control, maxwell2022entanglement, vrakking2022ion, ruberti2022quantum, he2023double, ishikawa2023control, ruberti2024bell}. Using weak XUV fields, significant experimental progress has been made \cite{akoury2007simplest,martin2007single,schoffler2011matter,fischer2013electron,waitz2016two,busto2022probing,koll2022experimental,nandi2024generation,Laurell2025,makos2025entanglement, koll2026entanglement}. 
    In particular, two key aspects of the quantum eraser concept have been demonstrated for molecular photoionization: first, the preparation and detection of an entangled ion-electron state \cite{martin2007single,shobeiry2024emission}; and, second, the entanglement-induced modification of the coherence accessible in one subsystem \cite{koll2022experimental}. 
    However, experimental studies of entanglement in strong-field ionization remain scarce \cite{eckart2023ultrafast} and so far the two key aspects of the quantum eraser have only been demonstrated in separate experiments. In this work, we combine both aspects of the quantum eraser, entanglement and interferometric coherence detection, within a single strong-field multiphoton ionization experiment. 

    To realize a photoelectron interferometer, we choose the process of strong-field multiphoton ionization of a diatomic molecule. Its inherent interference effects are best understood in the time domain \cite{becker2002above}: during each half-cycle of the linearly polarized laser field, an electron wavepacket tunnels from the molecule on alternating sides with respect to the origin. The free electron wavepackets are subsequently accelerated and dispersed by the laser electric field. Owing to their  coherence with one another, the wavepackets interfere. Wavepackets that are created in subsequent laser cycles lead to the above-threshold ionization (ATI) peaks \cite{agostini1979free} in the photoelectron spectrum. In addition, intra-cycle interferences \cite{paulus1998above, arbo2006} arise due to the interference of a wavepacket with itself or the one created in the following half-cycle. Notably, when such wavepackets scatter in the ionic field \cite{Corkum1993plasma}, rich interference structures known as photoelectron holography \cite{huismans2011time, bian2011subcycle} arise in the photoelectron momentum distribution. The holographic interferences represent a powerful imaging tool \cite{Faria2020it}, which is sensitive to properties of the scattering wavepacket \cite{meckel2014signatures} and the scattering potential \cite{Spanner2004}. Notably, holography arising from interfering wavepackets that originate from opposite sides of the molecule depends on the parity of the photoelectron wave function \cite{busuladvzic2008angle,kunitski2019double,Kang2020holographic}.

\begin{figure}[h!]
        \includegraphics[width=\linewidth]{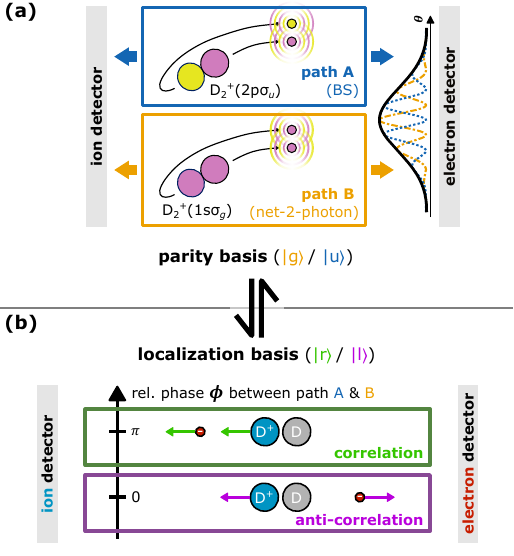}
        \caption{\small \justifying
        Schematic of the experiment. (a) During dissociative multiphoton ionization of ground-state D$_2$ molecules, an entangled pair of D$_2^+$ ion and photoelectron is created. This ion-photoelectron pair is entangled in parity and their state is denoted as $\ket{\psi_\mathrm{A+B}}$. For an even number of absorbed photons, electron and ion parities are either both odd (path A, bond-softening (BS) dissociation) or both even (path B, net-2-photon dissociation). To visualize the parity of ion and electron, the sign of the wavefunction is color-coded in yellow and pink. The D$_2^+$ ion dissociates and the resulting D$^+$ ion is detected in coincidence with the photoelectron. Two key observables are measured: the D$^+$ kinetic energy and a photoelectron interferogram. Both key observables depend on parity.
        On the electron detector side, the interference patterns for \textit{gerade} (orange, dot-dashed) and \textit{ungerade} (blue, dotted) photoelectron wave functions are schematically shown, as well as their incoherent sum (black). 
        (b) Illustration of the entanglement evidence: a basis transformation shows that the entangled state $\ket{\psi_\mathrm{A+B}}$ manifests itself in the predominant localization of photoelectron and ion depending on the relative phase $\phi$ between the paths A and B. For instance, $\phi=\pi$ refers to predominant localization of photoelectron and D$^+$ ion on the same side and thus predominant emission into the same direction (correlation). Hence, the entanglement in parity between D$_2^+$ ion and photoelectron results in entanglement in emission direction between D$^+$ ion and photoelectron after dissociation.}
        \label{fig:intro}
    \end{figure}

    The present work demonstrates that multiphoton interaction of molecules with an intense visible laser pulse creates entanglement between the photoelectron and the remaining molecular ion. Due to entanglement, measurements carried out on the ion sub-system may provide information on the photoelectron sub-system. The accessible information depends on the choice of measurement basis, as schematically shown in Fig.~\ref{fig:intro}. A measurement in the localization basis yields the emission direction of the dissociating molecular ion. Thus, which-way information identifying the emission side of the photoelectron becomes available. Consequently, photoelectron wavepackets emitted from opposite sides of the molecule become distinguishable and do not interfere with each other. A measurement in the parity basis, however, provides no information about the ion emission direction. Thus, photoelectron wavepackets emitted from opposite sides remain undistinguishable and interfere. Instead, the information about the parity would permit postselection of one of two complementary interference patterns associated with opposite parities. Our work complements studies of entanglement-induced decoherence in individual subsystems \cite{koll2022experimental, koll2026entanglement} by elucidating how coherence in an entangled state remains accessible through suitable joint measurements. We demonstrate how the choice of measurement basis for an entangled electron-ion system shapes the information accessible in ultrafast imaging, linking quantum-information concepts to strong-field photoelectron spectroscopy.
    
    In the following, we first establish the formation of the entangled photoelectron–ion state in multiphoton strong-field ionization and provide evidence for the entanglement. Next, it is shown how the presence of which-way information suppresses the holographic interference and how erasure of the which-way information restores the interference.
   
    
    An entangled electron-ion pair is created by the interaction of a neutral D$_2$ molecule with an intense (intensity $I \sim \SI{e14}{W/cm^2}$) femtosecond laser pulse at \SI{515}{nm} central wavelength, see methods. As illustrated in Fig.~\ref{fig:intro}(a) and detailed in the Supplementary Information (SI), after multiphoton ionization the dissociation of the D$_2^+$ molecular cation proceeds via two pathways, A and B, which are distinguished by the parity of the fragment ion. 
    Owing to parity conservation, the total parity of the bipartite system is even (odd) when the total number of absorbed photons is even (odd). Pathway A is defined by the ion being in the odd-parity $\mathrm{2p}\sigma_u$ state; therefore, the photoelectron must carry odd (even) parity for even (odd) photon orders. Conversely, in pathway B, the ion is in the even-parity $\mathrm{1s}\sigma_g$ state, requiring the photoelectron to be in an even (odd) state for even (odd) photon orders.
   Thus, for an even/odd total number of absorbed photons, the electron-ion wave function can be expressed as 
    \begin{equation}
        \ket{\psi_{\mathrm{A+B}}} =  \alpha\underbrace{\ket{\mathrm{e}_{u/g}}\ket{{\rm 2p}\sigma_u}}_{\text{path A}} + \beta e^{i \phi}\underbrace{\ket{\mathrm{e}_{g/u}} \ket{{\rm 1s}\sigma_g}}_{\text{path B}}, 
        \label{eq:psi_ug_basis}
    \end{equation}
    where $\ket{\mathrm{e}_{u/g}}$ represent photoelectron states with odd / even parity; $\alpha$ and $\beta$ represent real positive amplitudes, and $\phi$ is the relative phase of the two contributions. 
    As $\ket{\psi_{\mathrm{A+B}}}$ cannot be separated into a product state of two separate wave functions for electron and ion, $\ket{\psi_{\mathrm{A+B}}}$ describes an entangled state.
    If the amplitudes $\alpha = \beta$, $\ket{\psi_{\mathrm{A+B}}}$ corresponds to a maximally entangled Bell state. 

\begin{figure}[t]
        \centering
        \includegraphics[width=\linewidth]{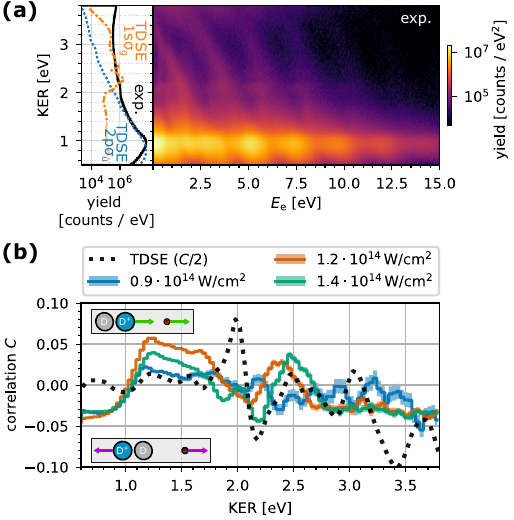}
        \caption{\small \justifying (a) Left: KER spectrum for experiment (solid black) and TDSE, where the calculated probability density is shown separately for the $\mathrm{1s}\sigma_g$ (dot-dashed orange) and the $\mathrm{2p}\sigma_u$ (dotted blue) state and scaled to the maximum in the experimental spectrum.  The experimental data is presented for a laser intensity of \SI{9e13}{W/cm^2} and pulse duration of \SI{35}{fs}, the TDSE for \SI{7.7e13}{W/cm^2} and \SI{52.5}{fs}. The TDSE laser parameters are chosen for best agreement with the experiment (see methods). Right: Measured joint energy spectrum showing the number of detected photoelectron-ion pairs as a function of KER and electron energy $E_\mathrm{e}$. 
        The experimental data is filtered to ions emitted along the laser polarization ($\theta_{\mathrm{D}^+}<\pi/8$ or $\theta_{\mathrm{D}^+}>7\pi/8$). 
        (b) Emission correlation parameter $C$ as a function of KER for experiment and TDSE (for an intensity of \SI{7.7e13}{W/cm^2}). For the experiment, $C$ is calculated according to Eq.~\ref{eq:correlation_exp}. The correlation $C$ is calculated for the TDSE by filtering the wavefunction to correlated and anti-correlated emission directions and integrating the probability density (\textit{i.e.} absolute squared of the wavefunction) over the 2D photoelectron momentum. Experimental data for different laser intensities are shown, as indicated in the figure legend. A moving average filter (width = three bins $\approx \SI{0.104}{\eV}$) is used to reduce noise in the experimental data. For comparability, the TDSE uses the same moving average filter. For good visibility of experiment and TDSE in the same plot, the shown TDSE correlation $C$ is downscaled by a factor of 2. The shaded area indicates the propagated standard deviation ($\sigma_C$), assuming Poissonian counting statistics ($\sigma_N=\sqrt{N}$ for $N$ coincidence events).}
        \label{fig:asymmetry}
    \end{figure}

    The parity of a particle's wavefunction cannot be directly measured. 
    However, the amplitudes $\alpha$ and $\beta$ of the two contributions of opposite parity to the wave function depend on the nuclear kinetic-energy release (KER). Thus, $\alpha$ and $\beta$ are tunable by selecting different KER regions.
    The KER ($=E_{\mathrm{D}^+}+E_{\mathrm{D}}$) results from propagation of the nuclear wavepacket on the potential energy curves associated with the electronic states of the molecular ion. Solving the time-dependent Schrödinger equation (TDSE) yields the KER distributions for each state, corresponding to the energy-dependent weights $\alpha^2(\mathrm{KER})$ and $\beta^2(\mathrm{KER})$ of Eq.~(\ref{eq:psi_ug_basis}). As seen in the left panel of Fig.~\ref{fig:asymmetry}(a), pathway A ($\mathrm{2p}\sigma_u$, dotted blue) dominates at $\mathrm{KER}\sim \SI{1}{eV}$, pathway B ($\mathrm{1s}\sigma_g$, dot-dashed orange) dominates at $\mathrm{KER}\sim \SI{3}{eV}$, and at $\mathrm{KER} \sim \SI{2}{\eV}$, pathways A and B contribute almost equally to the dissociation yield. 
    The photoelectron parity also cannot be measured directly, but the momentum distribution is dependent on parity, as schematically shown in Fig.~\ref{fig:intro}(a). 
    
    The correlated nature of the entangled state can be revealed through a measurement in a rotated basis.
    The new basis vectors, $\ket{l}=(\ket{g} - \ket{u})/\sqrt{2}$ and $\ket{r}=(\ket{g} + \ket{u})/\sqrt{2}$, describe a wavefunction localized predominantly on the left/right side in position space. Here, $\ket{g}$ ($\ket{u}$) refers to $\ket{\mathrm{e}_g}$ ($\ket{\mathrm{e}_u}$) in case of the photoelectron and to $\ket{{\rm 1s}\sigma_g}$ ($\ket{{\rm 2p}\sigma_u}$) in case of the bound electron in the molecular ion. The localization of the bound electron on the left/right nucleus causes the positive charge to be localized on the other nucleus. For clarity, we use capital letters when referring to the positive charge. Hence, in this $l/r$ basis 
    the entangled state of photoelectron and molecular ion becomes
    \begin{equation}
        \begin{aligned}
        \ket{\psi_{\mathrm{A+B}}} =
        [ (\alpha + \beta e^{i\phi})(&\ket{r, L} \pm \ket{l,R}) \\
        + (\beta e^{i\phi} - \alpha)(&\ket{r, R} \pm \ket{l,L}) ]  / 2 \; ,
        \end{aligned}
        \label{eq:psi_rL_basis}
    \end{equation}
    where the sign $+/-$ corresponds to even/odd number of absorbed photons. The two-particle state $\ket{r,L}$ ($\ket{l,R}$) represents a state where the photoelectron is predominantly localized on the right (left) side, while the positive charge in the molecular ion is predominantly localized on the left (right) atom. As discussed in the SI, the predominant localization of the wavefunction on one side transfers to a predominant emission towards the same side. In the case of $\ket{r,L}$ and $\ket{l,R}$, D$^+$ ion and the photoelectron are predominantly emitted into opposite hemispheres. For the states $\ket{r,R}$ and $\ket{l,L}$, ion and photoelectron are predominantly emitted into the same hemisphere. The basis transformation shows that the entanglement between ion and photoelectron parity results in a correlation in the $l/r$ basis, which is manifested as the correlated emission directions of the two particles, as illustrated in Fig.~\ref{fig:intro}. For a relative phase $\phi = \pi$, ion and electron are preferentially emitted into the same direction (\textquote{correlation}); for $\phi = 0$, ion and electron are preferentially emitted in opposite directions (\textquote{anti-correlation}). 
    
    As detailed in the SI, the correlated emission provides evidence of entanglement. In brief, for a well-defined (i.e. either even or odd) total parity, the correlated emission of ion and electron indicates a coherent superposition of pathways A and B [see Eq.~(\ref{eq:psi_ug_basis})]. For a fully incoherent classical mixture of the two-particle states in pathways A and B, i.e. in the absence of entanglement, the emission directions would be uncorrelated.

    To probe the emission direction correlation, the momentum vectors of the D$^+$ ions and photoelectrons, measured in coincidence, are analyzed. The measured data are presented as a joint-energy spectrum in Fig.~\ref{fig:asymmetry}.  The data exhibits diagonal lines, which reflect energy conservation of nuclei and electrons. Along each line, the total energy is constant and its value corresponds to the total number of photons absorbed from the field, either during ionization of the neutral or in the subsequent dissociation of the molecular ion. Hence, also the total parity is well-defined. The diagonal lines are most pronounced above a KER of \SI{1.5}{eV}, and somewhat blurred around $\mathrm{KER} = \SI{1}{eV}$, where the bond-softening process \cite{bucksbaum1990softening} dominates the dissociation yield. These observations place our experiment in an intermediate regime compared to previous work where strong energy correlation between photoelectron and ion was observed for $\SI{400}{nm}$ light \cite{wu2013electron} and little correlation was observed for intense $\SI{800}{nm}$ light \cite{lu2018high, He2020}. 

    The correlation in the emission directions of photoelectron and the D$^+$ ion are quantified by the emission direction correlation parameter 
    \begin{equation}
        C=\frac{N_+ - N_-}{N_+ + N_-} \; ,
        \label{eq:correlation_exp}
    \end{equation}
    where $N_+$ ($N_-$) is the number of electron-ion pairs emitted into same (opposite) directions, i.e., the angle $\theta$ between their momentum vectors is smaller than $\pi/6$ (larger than $5\pi/6$). These limits are chosen for best visibility of the correlation pattern. The observations remain qualitatively unchanged for other choices. $C>0$ corresponds to correlation and $C<0$ to anti-correlation. 
    

    Figure~\ref{fig:asymmetry}(b) presents a quantitative comparison of the measured and calculated emission correlation parameters as a function of KER. Above \SI{1.8}{\eV}, where the amplitudes of the contributions of path A and B become comparable, the emission correlation parameter exhibits a pronounced oscillation as a function of KER. The TDSE result shows a similar oscillatory behavior, which emphasizes the significance of the presented evidence for entanglement.   
    As shown in the SI, the emission correlation parameter is approximately $C(\ket{\psi_\mathrm{A+B}}) = -2 \alpha \beta \cos\phi$, \textit{cf.}~Eq.~(\ref{eq:asym_theory_entangl}). The expression is exact, if perfect spatial left/right localization in the $l$/$r$ basis is assumed - a condition that does not hold for the photoelectron, neither in the experiment nor in the TDSE. Consequently, even for $\alpha=\beta=1/\sqrt{2}$, the correlation remains smaller than $1$. The phase $\phi$ is accumulated during propagation on the potential energy curves, and therefore depends sensitively on the value of the KER. Importantly, the presence of the oscillation in $C$ around zero demonstrates the coherence of the two contributions in Eq.~(\ref{eq:psi_ug_basis}) to the total electron-ion wave function and thus provides evidence of entanglement. 

    Further, Fig.~\ref{fig:asymmetry}(b) shows measured values of $C$ for different laser intensities. With increasing intensity, the amplitude of the energy-dependent oscillations of $C$ increases, and their phase is shifted. This behavior is attributed to the non-linear laser-matter interaction in our experiment. At higher laser intensity, the relative contributions of paths A and B change, affecting the magnitude of $C$. The observed phase shift with varying intensity results from the laser-induced Stark shift, which distorts the potential-energy curves of the ${\rm 1s}\sigma_g$ and the ${\rm 2p}\sigma_u$ states. 
    We have performed additional TDSE calculations that confirm the intensity-dependent behavior of the amplitude and phase of the oscillations of $C(\mathrm{KER})$. In particular, because the phase of $C$ is strongly dependent on intensity, the magnitude of $C$ is significantly reduced when the measurement signal is integrated over the focal intensity distribution. This focal volume averaging has been neglected in the calculations, resulting in larger values of $C$ than observed in the experiment.
    Moreover, the reduced dimensionality of the TDSE may contribute to a larger $C$ as compared to the experiment.
    The appearance of emission correlation in the BS regime in the experimental data is discussed in the SI. 
    Next, the effect of the entanglement on the detected photoelectron is investigated.

    \begin{figure}
        \includegraphics[width=\linewidth]{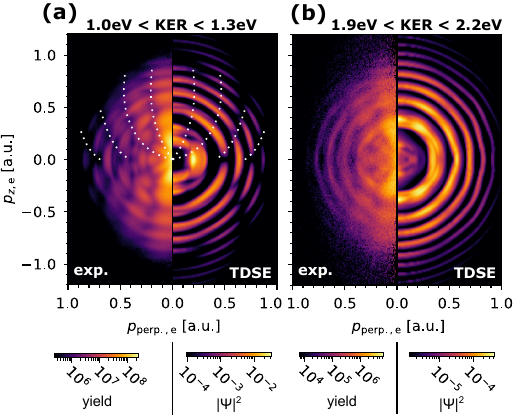}
        \caption{ \small \justifying 
            (a,b) Measured (left) and calculated (right) PMD in atomic units (a.u.) for photoelectrons emitted with D$^+$ ions of different KER. Panel (a) shows the PMD for dissociation dominated by path A.
            The minima of the holographic interference structures are visually highlighted by white, dotted lines. In (b), the PMD for dissociation with comparable contributions from paths A and B is shown. 
            The yield of the experimental data is given in units of $\mathrm{counts / a.u.}^3$. The experimental data corresponds to a laser intensity of \SI{9e13}{W/cm^2}, the TDSE results to \SI{7.7e13}{W/cm^2}. The pulse duration is \SI{35}{fs} in the experiment and \SI{52.5}{fs} in the TDSE. The TDSE laser parameters are chosen for best agreement with the experiment (see methods). The experimental data is filtered to ions emitted along the laser polarization ($\theta_{\mathrm{D}^+}<\pi/8$ or $\theta_{\mathrm{D}^+}>7\pi/8$). As the ion is modelled only in one dimension, the molecules are inherently oriented along the laser polarization in the TDSE. 
        }
        \label{fig:PMD}
    \end{figure}
    
    Figure~\ref{fig:PMD} presents measured and calculated photoelectron momentum distributions (PMD) side-by-side. Experiment and TDSE are in good agreement, though the calculated PMD exhibits stronger signals at large perpendicular momentum as a consequence of the reduced dimensionality of the calculations (2D). 
     The results presented in Fig.~\ref{fig:PMD}(a) and (b), have been filtered to ions with different KER values. 
     At both KERs, concentric rings are observed which correspond to ATI peaks, a consequence of interference between wavepackets created in subsequent laser cycles. The PMDs shown in Fig.~\ref{fig:PMD}(a) exhibit pronounced angular interference structures, i.e. the holographic interferences mentioned in the introduction. Strikingly, the holographic structures are strongly suppressed in Fig.~\ref{fig:PMD}(b), suggesting loss of photoelectron coherence between scattered wavepackets emitted upon tunnel ionization in subsequent half-cycles of the laser field.


    \begin{figure*}
        \centering
        \includegraphics[width=\textwidth]{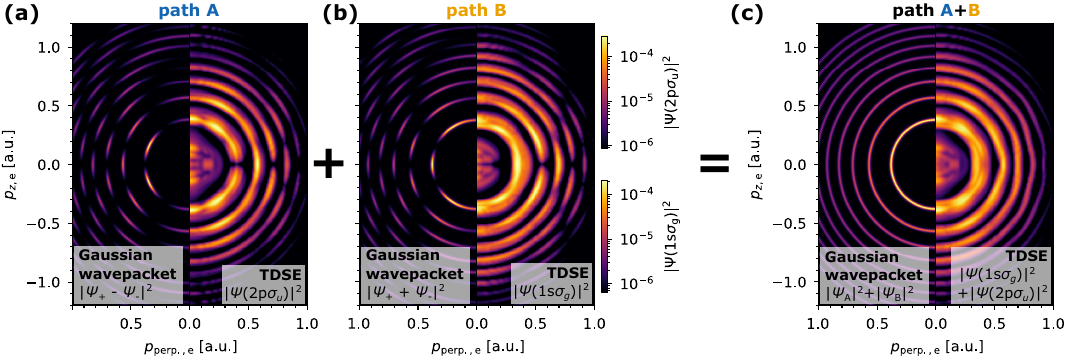}
        \caption{\small \justifying 
            Photoelectron momentum distributions for pathway A, pathway B, and their incoherent sum, as obtained from a simple interference model (left subplots) and the ab-initio TDSE calculations (right subplots) for an intensity of \SI{7.7e13}{W/cm^2}. Results are presented for pathway A (a), pathway B (b), and their incoherent sum (c), which is identical to the right half of Fig.~\ref{fig:PMD}(b).
        }
        \label{fig:holography}
    \end{figure*}

    To develop an intuitive understanding for the suppression of the interference pattern, we introduce a simplified model for photoelectron interference. In the following we use atomic units (a.u.), where $\hbar=e=m_e=4\pi\epsilon_0=1$. Our 2D model is based on two spatially separated, interfering wavepackets $\psi_+ +\psi_{-}$ in analogy to double-slit interference. We approximate the photoelectron wavepackets emitted by tunnel ionization in successive laser half-cycles by Gaussian wavepackets $\psi_{+}$ and $\psi_{-}$, launched at either side of the molecule, with 
    \begin{equation}
    \begin{aligned}
        &\psi_{\pm}(x,z,t)=\\&\quad\sum_n e^{-((z \mp d)^2+x^2)/\Delta r^2} e^{i(k_n\sqrt{(z \mp d)^2+x^2}-\omega_n (t \pm T/4))} \; .   
    \end{aligned}
    \end{equation}
    We use  $d=\SI{10}{\au}$, the approximate distance of the tunnel exits relative to the center of the molecule, and $\Delta r = \SI{150}{\au}$. The ATI peak structure is included by writing each wavepacket as a sum of spherical waves with energies $E_n=\omega_n = n \omega_{\SI{515}{nm}} + E_\mathrm{offset}$ and wavevectors $k_n=\sqrt{2E_n}$, where $n$ is the integer photon order and $E_\mathrm{offset}< \omega_{\SI{515}{nm}}$. The wavepackets $\psi_{+}$ and $\psi_{-}$ are centered on alternating sides of the molecule ($\pm d$) and are associated with initial times $\mp T/4$, which results from electric field flipping its sign every optical half-period $T/2$. Because of the time offsets, the alternating parities of even and odd photon orders appear naturally. The states $\ket{r}$ and $\ket{l}$ are approximated by $\psi_{+}$ and $\psi_{-}$, respectively. Similarly, $\psi_{+}+\psi_{-}$ corresponds to $\ket{e_{g/u}}$ of dissociation path B, $\psi_{+}-\psi_{-}$ corresponds to  $\ket{e_{u/g}}$ of dissociation path A.

    

    Figure~\ref{fig:holography}(a,b) presents calculated PMDs obtained from (left) the simple model and (right) the TDSE, for paths A and B. The interference features observed for the simple interference model qualitatively reproduce several features observed in the TDSE calculation. This agreement indicates that most of the holographic interference structures observed in TDSE and experiment can be assigned to the coherent superposition of wavepackets emitted from opposite sides of the molecule in successive laser half-cycles, as described by the simple model. 
    In the simple interference model, the only difference between pathways A and B is the opposite parity of the photoelectron. Clearly, this determines the position of interference maxima and minima. It can be seen, that the two interference patterns complement each other perfectly, meaning that the maxima obtained for one path overlap with the minima obtained for the other path (\textit{cf.} Fig.~\ref{fig:intro}). The TDSE calculations confirm this, as the interference pattern for pathway A and B observed in the TDSE results also complement each other, though not perfectly.  
        
    In Fig.~\ref{fig:holography}(c), the PMDs of pathways A and B are added incoherently. For the perfectly complementary results of the simple interference model, the interference patterns completely vanish. For the TDSE results, some modulations remain, but their contrast is drastically reduced as compared to the individual PMDs of pathway A and B. We conclude that, in the parity basis, the loss of the holographic interference pattern can be understood as the combination of contributions with opposite photoelectron parity to the PMD. 

    Alternatively, the loss of visibility of the interference pattern can be understood in terms of the availability of which-way information in the left/right basis. The primary component of our quantum eraser experiment is the holographic photoelectron interferogram. As illustrated by the simple model introduced above, it is best understood as a type of double-slit interference between the partial waves $\psi_{+}$ and $\psi_{-}$. The secondary component of the quantum eraser experiment is the measurement carried out on the entangled marker particle, i.e. the dissociating D$_2^+$ ion. If this measurement provides suitable which-way information, the partial waves do not interfere and the visibility of the interferogram vanishes. Indeed, as detailed in the SI, the measurement of the emission direction of the D$^+$ ion provides such which-way information. Our TDSE simulations show that for $\mathrm{KER}>\SI{1.8}{eV}$, the predominant D$^+$ emission direction oscillates with time at the laser frequency. Therefore, the measurement of the D$^+$ emission direction can be interpreted within the time-domain as knowledge of the optical phase at which ionization occurred, and thus as knowledge of the sign of the tunnel exit position. In terms of the simple interference model, the two partial wavepackets $\psi_{+}$ and $\psi_{-}$ are distinguished precisely by the sign of the tunnel exit positions. Thus, the ion emission direction represents which-way information: it identifies the ``slit", i.e. whether the entangled photoelectron is described by $\psi_{+}$ or $\psi_{-}$. Hence, the partial wavepackets are distinguishable and  will not interfere.

    Whether or not the measurement delivers which-way information that is relevant for the distinguishability of the contributions from $\ket{l}$ and $\ket{r}$ depends on the presence of an entangled state, i.e.~on the ion energy, see Fig.~\ref{fig:asymmetry}(b). 
    When selecting the KER region around \SI{1.15}{eV} [\textit{cf.} Fig.~\ref{fig:PMD}(a)], the total wave function contains contributions nearly exclusively from path A, i.e. we have $\alpha \approx 1$. Hence, the system is not in an entangled state but is described by the wavefunction $\ket{\psi_\mathrm{A}} = \ket{\mathrm{e}_{u/g}}\ket{{\rm 2p}\sigma_u}$ (path A), which is a separable product state of states of the photoelectron and the ion. 
    Because of the lack of entanglement, a measurement of the emission direction of the ion provides no information about the emission side of the photoelectron.  Thus $\ket{l}$ and $\ket{r}$ (or $\psi_+$ and $\psi_{-}$ in the simple model) add coherently, and holographic interference is visible in the PMD. 
    In Fig.~\ref{fig:PMD}(b), the photoelectron data is filtered to the KER range, where the photoelectron and the ion are in an entangled state with nearly equal contributions from paths A and B ($\alpha \approx \beta$). Now, the ion emission direction provides relevant which-way information. This causes the contributions from the photoelectron states $\ket{l}$ and $\ket{r}$ to enter the PMD in Fig.~\ref{fig:PMD}(b) only as an incoherent sum, which results in suppression of the holographic interference pattern. 

    
    The eponymous key observation of the quantum eraser lies in the restoration of the interference pattern by post-selecting on the measured state of the marker particle. How can we restore the coherence between $\ket{l}$ and $\ket{r}$ in the present experiment?


    In order to observe the coherence and erase the which-way information, the measurement of the PMDs needs to be carried out in the gerade/ungerade basis of the ion. The eigenstates of this basis mask the emission direction and, hence, the ion cannot carry which-way information.    
    Using the electron-ion TDSE solutions, we can directly filter the PMD to either the $\mathrm{1s}\sigma_g$ or the $\mathrm{2p}\sigma_u$ state of the ion and recover the holographic interference, as shown in Fig.~\ref{fig:holography}. 
    Though we cannot directly measure the parity of the $\mathrm{D}_2^+$ ion in the present experiment, we can select the measured data for states with either \textit{gerade} or \textit{ungerade} parity by choosing suitable KER regions, where one of the contributions dominates. In these cases, the holographic interference is clearly observed, as for example in Fig.~\ref{fig:PMD}(a), where \textit{ungerade} parity of the ion dominates.  

    We have demonstrated the preparation of an entangled state consisting of a photoelectron and a dissociating molecular ion by strong-field dissociative ionization of D$_2$.
    In our experiment, the emission direction of the ion after dissociation may carry which-way information about the photoelectron. If the parity of the ion is not detected, the holographic interference pattern in the PMD is lost. We demonstrate the restoration of the holographic pattern within the TDSE solutions, namely by post-selection for the parity of the ion. Experimental detection of the ion parity, e.g., by using an intense probe pulse in the XUV \cite{spathe2025laser}, would be an ambitious goal for future studies. Our work exemplifies how fundamental concepts of quantum information science can be studied in the context of ultrafast photoionization using coincidence measurements of ions and electrons. Strong-field and attosecond science provides a powerful toolbox for the ultrafast preparation of electron-ion pairs in entangled states and the manipulation of ultrafast molecular dynamics. Emerging attosecond quantum-information science may shed light on entanglement and decoherence effects during the first steps of unimolecular reactions.
    
    \begin{acknowledgments}
        We thank Th. Weber and F. Ronneberger for their technical support. This project has been funded by the Deutsche Forschungsgemeinschaft (DFG, German Science Foundation) under the Emmy Noether programme project No. 437321733 and under the project No. 498967973. The present work was also supported by the Collaborative Research Center 1375 \textquote{Nonlinear optics down to atomic scales} (NOA) under projects B1 and A7 (project No. 398816777). Numerical simulations were mostly performed on compute clusters at the Leibniz University Hannover, which were funded by the Deutsche Forschungsgemeinschaft (DFG, German Research Foundation) under project numbers 411116428, 424969120, and 563020064. Further financial support has been provided by the profile line LIGHT by the Friedrich Schiller University and by the Max Planck School of Photonics. We thank M.J.J. Vrakking, J.M. Dahlström, and R. Keil for fruitful discussions. 
    \end{acknowledgments}

  \section{Author contributions}
  S.H, J.S., and M.K. conducted the experiment and S.H and M.K. analyzed it. S.H., P.W., M.G., S.M., D.B.M., M.L., and M.K. performed analytical or numerical calculations and evaluated them. S.H. and M.K. wrote the manuscript. All authors discussed the results and revised the manuscript.   
  
    \section{Data availability}
    The processed experimental data and the numerical data underlying all TDSE plots are publicly available at \cite{hell2026data}. These datasets, together with the accompanying analysis code, enable reproduction of the reported figures. The unprocessed experimental data and full raw TDSE outputs are not included in the repository, as their interpretation and processing require specialist expertise in coincidence spectroscopy and TDSE simulations, respectively. The raw data are available from the authors upon reasonable request.  

    \section{Code Availability}
    The TDSE code employed for the theoretical calculations presented in Figs. \ref{fig:asymmetry}, \ref{fig:PMD}, \ref{fig:holography}, \ref{fig:KER_all_intensities}, and \ref{fig:asym_all_intensities} requires several hundred thousand CPU hours on high-performance computing facilities, such as the system acknowledged here. Its use also requires specialized expertise and guidance from the code developers. Access to the code, together with the necessary support for its use, can be provided by the authors upon reasonable request.

  \section{Methods}

    The experiment is carried out using $\sim\SI{35}{fs}$ (FWHM) laser pulses, centered at \SI{1030}{nm} and provided at a repetition rate of \SI{100}{kHz} by a commercial femtosecond laser system (Active Fiber Systems). Next, the laser pulses are frequency-doubled in a \SI{300}{\micro m} thick beta barium borate (BBO) crystal. Subsequently, we split off the residual \SI{1030}{nm} light, clean the polarization of the generated \SI{515}{nm} pulses and compensate any chirp acquired in the BBO or in transmissive optics. The resulting $\sim \SI{35}{fs}$ (FWHM) pulses are then focused by a back-focusing spherical mirror with \SI{75}{mm} focal length into a cold D\textsubscript{2} gas jet inside a COLd Target Recoil Ion Momentum Spectrometer (COLTRIMS) \cite{ullrich2003recoil}. We choose D\textsubscript{2} to avoid contamination of the signal with fragments emitted from background H$_2$ gas. 
    Using time and position sensitive detectors on either side of the COLTRIMS, electrons and ions are detected in coincidence. Limiting the count rate to $<\SI{0.2}{events/shot}$  allows to distinguish pure ionization and dissociative ionization by either detecting D$_2^+$ or D$^+$ in coincidence with an electron. The neutral D fragment from dissociative ionization is not detected, but its momentum is reconstructed based on momentum conservation.
    Since the ionization of background H$_2$ gas overlaps with low energy D$^+$ ions, the presented experimental data is limited to $\mathrm{KER}>\SI{0.5}{eV}$. 
    
    The intensity of the visible laser field in the focus is estimated based on the high-energy cutoff of the photoelectron spectra detected in coincidence with D$_2^+$. The experiment has been conducted for four different intensities ranging from $I_{\SI{515}{nm}} \approx \SI{9e13}{W/cm^2}$ ($U_\mathrm{P}\approx\SI{2.2}{\eV}$) up to $I_{\SI{515}{nm}} \approx \SI{1.4e14}{W/cm^2}$ ($U_\mathrm{P}\approx\SI{3.5}{\eV}$).
    

    To simulate the experiment, we use a non-Born-Oppenheimer grid-based model for the dissociative ionization of the D$_2$ molecule \cite{winter2026free}. The dynamics of the molecule comprises three parts: the nuclei, a bound electron, and an active electron that can be freed by laser-induced ionization.
    The bound electron is treated in a dipole-coupled two-level system, where the two levels correspond to the two lowest electronic states ($\mathrm{1s}\sigma_g$ and $\mathrm{2p}\sigma_u$) when the system is ionized or the lowest dipole-coupled states of the neutral molecule ($X^1\Sigma_g^+$ and $B^1\Sigma_u^+$) when the active electron is in its lowest possible state.
    The active electron is described in two dimensions [$\mathbf{r} = (x,z)^T$] and the motion of the nuclei in one dimension with the coordinate being the internuclear distance $R$.
    Rotation and center-of-mass motion is not considered.
    This results in a two-component three-dimensional wavefunction $\Psi(t) = (\psi_g(\mathbf{r},R,t), \psi_u(\mathbf{r},R,t))^T$ where $g$/$u$ denotes the \textit{gerade}/\textit{ungerade} symmetry of the bound electron.
    The active electron and the core interact via a soft-core potential.
    Parameters are optimized to fit literature values for the ionization potentials for both symmetries ($g$/$u$).
    This results in the time-dependent Schr\"odinger equation
    \begin{equation}
    \begin{aligned}
        &{\rm i}\partial_t \begin{pmatrix}
            \psi_g(\mathbf{r},R,t)\\
            \psi_u(\mathbf{r},R,t)
        \end{pmatrix}
        \\=
        &\begin{pmatrix}
            \hat{\mathbf{H}}_g & -\mathbf{d}(\mathbf{r},R)\cdot \bm{\mathcal{E}}(t) \\
            -\mathbf{d}(\mathbf{r},R)\cdot\bm{\mathcal{E}}(t) & \hat{\mathbf{H}}_u
        \end{pmatrix}
        \begin{pmatrix}
            \psi_g(\mathbf{r},R,t)\\
            \psi_u(\mathbf{r},R,t)
        \end{pmatrix}\, ,
    \end{aligned}
    \end{equation}
    where $\mathbf{d}$ is the transition dipole moment and $\bm{\mathcal{E}}$ is the electric field. The diagonal terms are given by
    \begin{equation}
        \hat{\mathbf{H}}_{g/u} = \frac{P^2}{2 \mu} + \frac{\mathbf{p}^2}{2} + V_{g/u}^{\rm
            int}(\mathbf{r},R)+V_{g/u}^{\rm ion}(R)\, ,
    \end{equation}
    where $\mu$ is the reduced mass of the nuclei, $V_{g/u}^{\rm ion}$ are the Born-Oppenheimer potential curves of the ion's
    lowest two electronic states and $V_{g/u}^{\rm int}$ is the soft-core interaction
    potential. A $384\times 384\times 256$ grid ($\Delta x=\Delta z=0.4\,$a.u., $\Delta R=0.05\,$a.u.) is used when the active electron is near the nuclei, and a zero-padded $1536\times 1536\times 1024$ grid for large separations (cf.~\cite{winter2026free}), giving a momentum resolution of $\Delta p_x=\Delta p_z\approx 0.01\,$a.u. for the PMD.

    To calculate the emission direction correlation $C$ from the TDSE results, the \textit{gerade} and \textit{ungerade} components of the wavefunction are first transformed to the localized basis, $\psi_{r/l}=(\psi_{\mathrm{1s}\sigma_g}(\mathbf{p}_\mathrm{e},R)\pm\psi_{\mathrm{2p}\sigma_u}(\mathbf{p}_\mathrm{e},R))/\sqrt{2}$, corresponding to the two opposite emission directions of the D$^{+}$ fragment at the end of the calculation. Next, these two wavefunctions are Fourier-transformed with respect to the internuclear distance $R$, yielding the wavefunction as a function of the photoelectron momentum $\mathbf{p}_\mathrm{e}=(p_x,p_z)$ and nuclear momentum $P$. For each KER ($=P^2/(2\mu)$), the probability densities $|\psi_{r/l}(\mathbf{p}_\mathrm{e},P)|^2$ are integrated over the two-dimensional photoelectron momentum distribution within the angular ranges corresponding to same- and opposite-direction electron–ion emission. Specifically, $N_+$ is obtained by integrating over photoelectron momenta for which the angle $\theta$ between the electron and D$^{+}$-ion emission directions satisfies $\theta<\pi/6$, whereas $N_-$ is obtained for $\theta>5\pi/6$, analogous to the analysis of the experimental data. Since the nuclear motion is restricted to the laser-polarization axis $z$ in the TDSE, $\theta$ is determined by the photoelectron emission angle in the $(p_x,p_z)$ plane. The emission direction correlation $C$ is then calculated by inserting $N_+$ and $N_-$ in  Eq.~\ref{eq:correlation_exp}.

    To obtain the experimental emission-direction correlation $C$ as a function of KER [\textit{cf.} Fig.~\ref{fig:asymmetry}(b)], coincidence events are first filtered to ions emitted along the laser polarization axis ($\theta_{\mathrm{D}^+}<\pi/8$ or $\theta_{\mathrm{D}^+}>7\pi/8$). The selected events are then sorted into two KER-resolved histograms, $N_+(\mathrm{KER})$ and $N_-(\mathrm{KER})$, corresponding to electron–ion pairs emitted in the same ($\theta<\pi/6$) and opposite ($\theta>5\pi/6$) directions, respectively. Here, $\theta$ denotes their relative emission angle. The experimental $C(\mathrm{KER})$ is then obtained by inserting $N_+(\mathrm{KER})$ and $N_-(\mathrm{KER})$ in Eq.~\ref{eq:correlation_exp}. While the histograms $N_+(\mathrm{KER})$ and $N_-(\mathrm{KER})$ include photoelectrons of all measured kinetic energies, limiting the ion emission angle and the relative emission angle of ion and photoelectron excludes electrons emitted approximately perpendicular to the laser polarization axis.
   
    Laser intensity and pulse duration affect the emission-direction correlation $C$, the photoelectron momentum distribution (PMD), and the relative contributions of dissociation paths A ($\mathrm{2p}\sigma_u$) and B ($\mathrm{1s}\sigma_g$). Comparing experiment and theory therefore requires accounting for uncertainties in the experimental laser parameters and the absence of focal-volume averaging in the TDSE calculations.

    To guide the choice of simulation parameters, we first scanned the peak intensity and pulse duration using a reduced 1D (ion-only), two-level TDSE model. All pulse durations quoted in this work refer to the FWHM of the intensity envelope ($\bm{\mathcal{E}}^2$). The BS peak shifted towards lower energies with increasing intensity and pulse duration, with mean slopes of \( (\num[uncertainty-mode = separate]{-0.3 +- 0.1})\,\unit{\eV / (10^{14}\, W/cm^2)}\) and \( (\num[uncertainty-mode = separate]{-0.012 +- 0.003})\,\unit{\eV/ fs}\), respectively. These values represent the mean and standard deviation of the slopes over the investigated ranges of \SIrange{5.6e13}{1e14}{W/cm^2} and \SIrange{35}{70}{fs}, reflecting their dependence on both parameters. A shift in the BS peak position changes the relative contributions of paths A and B at a given KER, thereby modifying the combined PMD because the two paths exhibit complementary holographic pattern (compare Fig.~\ref{fig:holography}). A complementary scan using a 2D (electron-only), fixed-nuclei TDSE model showed that the above-threshold ionization (ATI) peaks in the photoelectron energy spectrum shift approximately according to the intensity-dependent ponderomotive potential.

    Guided by these scans, we selected a small set of intensity and pulse-duration combinations for high-resolution calculations with the full 3D TDSE model. As observed already in the 1D (ion-only) calculations, the relative contribution of path B increases with increasing laser intensity (see Appendix~\ref{sec:KER_spectra_diff_int_diff_pulse_dur}), which affects the PMD as mentioned above. Moreover, the full 3D calculations show that with increasing pulse duration the ATI peaks of path B shift by $\approx\SI{3.66}{m\eV/fs}$ and the ATI peaks of path A by $\approx\SI{-12.1}{m\eV/fs}$ (the ATI peaks of the bound vibrational states, \textit{i.e.} photoelectrons of non-dissociating D$_2^+$, are not affected by laser pulse duration). Together, these dependencies constrain the parameter combinations for which the ATI peaks of both dissociation paths overlap and their dissociation yields are comparable in the range $\SI{1.9}{\eV} < \mathrm{KER} < \SI{2.2}{\eV}$, as required for agreement with the experimental PMD in Fig.3(b). The best agreement with experiment was obtained at a peak intensity of \SI{7.7e13}{W/cm^2} and a pulse duration of \SI{52.5}{fs} for a $\sin^2$ pulse envelope; these parameters were therefore used for the TDSE results presented here. In the light of uncertainties in the experimental laser parameters and the absence of focal volume averaging in the TDSE, the discrepancy of 15\% between experimental intensity and used in the TDSE calculations is considered reasonable.

    Large Language Models (ChatGPT-5, ChatGPT-6 and Gemini 3) were employed to improve the readability and conciseness of the original manuscript text. All scientific content and conclusions were generated by the authors, who maintain full accountability for the final text.

    \bibliography{apssamp}

@PREAMBLE{
 "\providecommand{\noopsort}[1]{}" 
 # "\providecommand{\singleletter}[1]{#1}%" 
}

@article{agostini1979free,
  title={Free-free transitions following six-photon ionization of xenon atoms},
  author={Agostini, Pierre and Fabre, F and Mainfray, G{\'e}rard and Petite, Guillaume and Rahman, N Ko},
  journal={Phys. Rev. Lett.},
  volume={42},
  number={17},
  pages={1127},
  year={1979},
  publisher={APS}
}

@article{paulus1998above,
  title={Above-threshold ionization by an elliptically polarized field: quantum tunneling interferences and classical dodging},
  author={Paulus, G G and Zacher, F and Walther, H and Lohr, A and Becker, W and Kleber, M},
  journal={Phys. Rev. Lett.},
  volume={80},
  number={3},
  pages={484},
  year={1998},
  publisher={APS}
}

@article{winter2026free,
  title={Free-electron vortices in photoelectron momentum distributions from molecules with coupled electronic-nuclear motion},
  author={Winter, Paul and Lein, Manfred},
  journal={Phys. Rev. A},
  volume={},
  number={},
  pages={},
  year={2026},
  publisher={APS},
  doi = {10.1103/s4cm-rhrm},
  url = {https://link.aps.org/doi/10.1103/s4cm-rhrm}
}

@dataset{hell2026data,
    author = {Hell, Sebastian and
              Winter, Paul and
              Späthe, Julian and
              Paulus, Gerhard G. and
              Lein, Manfred and
              Kübel, Matthias},
    title = {Research data for: Entanglement and photoelectron
             holography in dissociative photoionization:
             molecular quantum eraser},
    year = {2026},
    publisher = {Zenodo},
    doi = {10.5281/zenodo.23314216}
}

@article{yin1992asymmetric,
  title = {Asymmetric photoelectron angular distributions from interfering photoionization processes},
  author = {Yin, Yi-Yian and Chen, Ce and Elliott, D. S. and Smith, A. V.},
  journal = {Phys. Rev. Lett.},
  volume = {69},
  issue = {16},
  pages = {2353--2356},
  numpages = {0},
  year = {1992},
  month = {Oct},
  publisher = {American Physical Society},
  doi = {10.1103/PhysRevLett.69.2353},
  url = {https://link.aps.org/doi/10.1103/PhysRevLett.69.2353}
}

@article{lafosse2003molecular,
  title={Molecular frame photoelectron angular distributions in dissociative photoionization of H2 in the region of the Q1 and Q2 doubly excited states},
  author={Lafosse, A and Lebech, M and Brenot, JC and Guyon, PM and Spielberger, L and Jagutzki, O and Houver, JC and Dowek, D},
  journal={J. Phys. B},
  volume={36},
  number={23},
  pages={4683--4702},
  year={2003}
}

@article{brennecke2020gouy,
  title={Gouy’s phase anomaly in electron waves produced by strong-field ionization},
  author={Brennecke, Simon and Eicke, Nicolas and Lein, Manfred},
  journal={Phys. Rev. Lett.},
  volume={124},
  number={15},
  pages={153202},
  year={2020},
  publisher={APS}
}

@article{arbo2006,
  title = {Time double-slit interferences in strong-field tunneling ionization},
  author = {Arb\'o, Diego G and Persson, Emil and Burgd\"orfer, Joachim},
  journal = {Phys. Rev. A},
  volume = {74},
  issue = {6},
  pages = {063407},
  numpages = {6},
  year = {2006},
  month = {Dec},
  publisher = {American Physical Society},
  doi = {10.1103/PhysRevA.74.063407},
  url = {https://link.aps.org/doi/10.1103/PhysRevA.74.063407}
}

@article{giusti1990above,
  title={Above-threshold dissociation of {H}$_2^+$ in intense laser fields},
  author={Giusti-Suzor, Annick and He, Xin and Atabek, Osman and Mies, Frederic H},
  journal={Phys. Rev. Lett.},
  volume={64},
  number={5},
  pages={515},
  year={1990},
  publisher={APS}
}

@article{Scully1982quantum,
  title = {Quantum eraser: A proposed photon correlation experiment concerning observation and \textquote{delayed choice} in quantum mechanics},
  author = {Scully, Marlan O. and Dr\"uhl, Kai},
  journal = {Phys. Rev. A},
  volume = {25},
  issue = {4},
  pages = {2208--2213},
  numpages = {0},
  year = {1982},
  month = {Apr},
  publisher = {American Physical Society},
  doi = {10.1103/PhysRevA.25.2208},
  url = {https://link.aps.org/doi/10.1103/PhysRevA.25.2208}
}

@article{Kang2020holographic,
  title = {Holographic detection of parity in atomic and molecular orbitals},
  author = {Kang, HuiPeng and Maxwell, Andrew S. and Trabert, Daniel and Lai, XuanYang and Eckart, Sebastian and Kunitski, Maksim and Sch\"offler, Markus and Jahnke, Till and Bian, XueBin and D\"orner, Reinhard and Faria, Carla Figueira de Morisson},
  journal = {Phys. Rev. A},
  volume = {102},
  issue = {1},
  pages = {013109},
  numpages = {7},
  year = {2020},
  month = {Jul},
  publisher = {American Physical Society},
  doi = {10.1103/PhysRevA.102.013109},
  url = {https://link.aps.org/doi/10.1103/PhysRevA.102.013109}
}

@article{Faria2020it,
doi = {10.1088/1361-6633/ab5c91},
url = {https://doi.org/10.1088/1361-6633/ab5c91},
year = {2020},
month = {jan},
publisher = {IOP Publishing},
volume = {83},
number = {3},
pages = {034401},
author = {Figueira de Morisson Faria, C and Maxwell, A S},
title = {It is all about phases: ultrafast holographic photoelectron imaging},
journal = {Rep. Prog. Phys.}
}

@article{makos2025entanglement,
  title={Entanglement in photoionisation reveals the effect of ionic coupling in attosecond time delays},
  author={Makos, Ioannis and Busto, David and Benda, Jakub and Ertel, Dominik and Merzuk, Barbara and Steiner, Benjamin and Frassetto, Fabio and Poletto, Luca and Schr{\"o}ter, Claus Dieter and Pfeifer, Thomas and others},
  journal={Nat. Commun.},
  volume={16},
  number={1},
  pages={8554},
  year={2025},
  publisher={Nature Publishing Group UK London}
}

@article{shobeiry2024emission,
  title={Emission control of entangled electrons in photoionisation of a hydrogen molecule},
  author={Shobeiry, Farshad and Fross, Patrick and Srinivas, Hemkumar and Pfeifer, Thomas and Moshammer, Robert and Harth, Anne},
  journal={Sci. Rep.},
  volume={14},
  number={1},
  pages={19630},
  year={2024},
  publisher={Nature Publishing Group UK London}
}

@article{nandi2024generation,
  title={Generation of entanglement using a short-wavelength seeded free-electron laser},
  author={Nandi, Saikat and Stenquist, Axel and Papoulia, Asimina and Olofsson, Edvin and Badano, Laura and Bertolino, Mattias and Busto, David and Callegari, Carlo and Carlstr{\"o}m, Stefanos and Danailov, Miltcho B and others},
  journal={Sci. Adv.},
  volume={10},
  number={16},
  pages={eado0668},
  year={2024},
  publisher={American Association for the Advancement of Science}
}

@article{koll2022experimental,
  title={Experimental control of quantum-mechanical entanglement in an attosecond pump-probe experiment},
  author={Koll, Lisa-Marie and Maikowski, Laura and Drescher, Lorenz and Witting, Tobias and Vrakking, Marc J J},
  journal={Phys. Rev. Lett.},
  volume={128},
  number={4},
  pages={043201},
  year={2022},
  publisher={APS}
}

@article{busto2022probing,
  title={Probing electronic decoherence with high-resolution attosecond photoelectron interferometry},
  author={Busto, David and Laurell, Hugo and Finkelstein-Shapiro, Daniel and Alexandridi, Christiana and Isinger, Marcus and Nandi, Saikat and Squibb, Richard J and Turconi, Margherita and Zhong, Shiyang and Arnold, Cord L and others},
  journal={Eur. Phys. J. D},
  volume={76},
  number={7},
  pages={112},
  year={2022},
  publisher={Springer}
}

@article{waitz2016two,
  title={Two-particle interference of electron pairs on a molecular level},
  author={Waitz, M and Metz, D and Lower, J and Schober, C and Keiling, M and Pitzer, Martin and Mertens, K and Martins, M and Viefhaus, J and Klumpp, S and others},
  journal={Phys. Rev. Lett.},
  volume={117},
  number={8},
  pages={083002},
  year={2016},
  publisher={APS}
}

@article{fischer2013electron,
  title={Electron Localization Involving Doubly Excited States in Broadband Extreme Ultraviolet Ionization of {H}$_2$},
  author={Fischer, Andreas and Sperl, Alexander and C{\"o}rlin, Philipp and Sch{\"o}nwald, Michael and Rietz, Helga and Palacios, Alicia and Gonz{\'a}lez-Castrillo, Alberto and Mart{\'\i}n, Fernando and Pfeifer, Thomas and Ullrich, Joachim and others},
  journal={Phys. Rev. Lett.},
  volume={110},
  number={21},
  pages={213002},
  year={2013},
  publisher={APS}
}

@article{schoffler2011matter,
  title={Matter wave optics perspective at molecular photoionization: {K}-shell photoionization and Auger decay of {N}$_2$},
  author={Sch{\"o}ffler, MS and Jahnke, T and Titze, J and Petridis, N and Cole, K and Schmidt, L Ph H and Czasch, A and Jagutzki, O and Williams, JB and Cocke, CL and others},
  journal={New J. Phys.},
  volume={13},
  number={9},
  pages={095013},
  year={2011},
  publisher={IOP Publishing}
}

@article{martin2007single,
  title={Single photon-induced symmetry breaking of {H}$_2$ dissociation},
  author={Mart{\'\i}n, F and Fern{\'a}ndez, J and Havermeier, T and Foucar, L and Weber, Th and Kreidi, K and Schoffler, M and Schmidt, L and Jahnke, T and Jagutzki, O and others},
  journal={Science},
  volume={315},
  number={5812},
  pages={629--633},
  year={2007},
  publisher={American Association for the Advancement of Science}
}

@article{akoury2007simplest,
  title={The simplest double slit: interference and entanglement in double photoionization of {H}$_2$},
  author = {D Akoury  and K Kreidi  and T Jahnke  and Th Weber  and A Staudte  and M Schöffler  and N Neumann  and J Titze  and L Ph H Schmidt  and A Czasch  and O Jagutzki  and R A Costa Fraga  and R E Grisenti  and R Díez Muiño  and N A Cherepkov  and S K Semenov  and P. Ranitovic  and C L Cocke  and T Osipov  and H Adaniya  and J C Thompson  and M H Prior  and A Belkacem  and A L Landers  and H Schmidt-Böcking  and R Dörner},
  journal={Science},
  volume={318},
  number={5852},
  pages={949--952},
  year={2007},
  publisher={American Association for the Advancement of Science}
}

@article{ruberti2024bell,
  title={Bell test of quantum entanglement in attosecond photoionization},
  author={Ruberti, Marco and Averbukh, Vitali and Mintert, Florian},
  journal={Phys. Rev. X},
  volume={14},
  number={4},
  pages={041042},
  year={2024},
  publisher={APS}
}

@article{busuladvzic2008angle,
  title={Angle-Resolved High-Order Above-Threshold Ionization of a Molecule: Sensitive Tool for Molecular Characterization},
  author={Busulad{\v{z}}i{\'c}, M and Gazibegovi{\'c}-Busulad{\v{z}}i{\'c}, A and Milo{\v{s}}evi{\'c}, D B and Becker, W},
  journal={Phys. Rev. Lett.},
  volume={100},
  number={20},
  pages={203003},
  year={2008},
  publisher={APS}
}

@article{ishikawa2023control,
  title={Control of ion-photoelectron entanglement and coherence via Rabi oscillations},
  author={Ishikawa, Kenichi L and Prince, Kevin C and Ueda, Kiyoshi},
  journal={J. Phys. Chem. A},
  volume={127},
  number={50},
  pages={10638--10646},
  year={2023},
  publisher={ACS Publications}
}

@article{he2023double,
  title={Double-slit interference in the ion dynamics of dissociative photoionization},
  author={He, Pei-Lun and Hatsagortsyan, Karen Z and Keitel, Christoph H},
  journal={Phys. Rev. Lett.},
  volume={131},
  number={1},
  pages={013201},
  year={2023},
  publisher={APS}
}

@article{vrakking2022ion,
  title={Ion-photoelectron entanglement in photoionization with chirped laser pulses},
  author={Vrakking, Marc J J},
  journal={J. Phys. B},
  volume={55},
  number={13},
  pages={134001},
  year={2022},
  publisher={IOP Publishing}
}

@article{vrakking2021control,
  title={Control of attosecond entanglement and coherence},
  author={Vrakking, Marc J J},
  journal={Phys. Rev. Lett.},
  volume={126},
  number={11},
  pages={113203},
  year={2021},
  publisher={APS}
}

@article{nishi2019entanglement,
  title={Entanglement and coherence in photoionization of {H}$_2$ by an ultrashort {XUV} laser pulse},
  author={Nishi, Takanori and L{\"o}tstedt, Erik and Yamanouchi, Kaoru},
  journal={Phys. Rev. A},
  volume={100},
  number={1},
  pages={013421},
  year={2019},
  publisher={APS}
}

@article{kunitski2019double,
  title={Double-slit photoelectron interference in strong-field ionization of the neon dimer},
  author={Kunitski, Maksim and Eicke, Nicolas and Huber, Pia and K{\"o}hler, Jonas and Zeller, Stefan and Voigtsberger, J{\"o}rg and Schlott, Nikolai and Henrichs, Kevin and Sann, Hendrik and Trinter, Florian and others},
  journal={Nat. Commun.},
  volume={10},
  number={1},
  pages={1},
  year={2019},
  publisher={Nature Publishing Group UK London}
}

@article{majorosi2017quantum,
  title={Quantum entanglement in strong-field ionization},
  author={Majorosi, Szil{\'a}rd and Benedict, Mih{\'a}ly G and Czirj{\'a}k, Attila},
  journal={Phys. Rev. A},
  volume={96},
  number={4},
  pages={043412},
  year={2017},
  publisher={APS}
}

@article{rohringer2009multichannel,
  title={Multichannel coherence in strong-field ionization},
  author={Rohringer, Nina and Santra, Robin},
  journal={Phys. Rev. A},
  volume={79},
  number={5},
  pages={053402},
  year={2009},
  publisher={APS}
}

@article{spanner2007entanglement,
  title={Entanglement and timing-based mechanisms in the coherent control of scattering processes},
  author={Spanner, Michael and Brumer, Paul},
  journal={Phys. Rev. A},
  volume={76},
  number={1},
  pages={013408},
  year={2007},
  publisher={APS}
}

@article{fedorov2004packet,
  title={Packet narrowing and quantum entanglement in photoionization and photodissociation},
  author={Fedorov, M V and Efremov, M A and Kazakov, A E and Chan, K W and Law, C K and Eberly, J H},
  journal={Phys. Rev. A},
  volume={69},
  number={5},
  pages={052117},
  year={2004},
  publisher={APS}
}

@article{eckart2023ultrafast,
  title={Ultrafast preparation and detection of entangled atoms},
  author={Eckart, Sebastian and Trabert, Daniel and Rist, Jonas and Geyer, Angelina and Schmidt, Lothar Ph H and Fehre, Kilian and Kunitski, Maksim},
  journal={Sci. Adv.},
  volume={9},
  number={36},
  pages={eabq8227},
  year={2023},
  publisher={American Association for the Advancement of Science}
}

@article{maxwell2022entanglement,
  title={Entanglement of orbital angular momentum in non-sequential double ionization},
  author={Maxwell, Andrew S and Madsen, Lars Bojer and Lewenstein, Maciej},
  journal={Nat. Commun.},
  volume={13},
  number={1},
  pages={4706},
  year={2022},
  publisher={Nature Publishing Group UK London}
}

@article{kremer2009electron,
  title={Electron Localization in Molecular Fragmentation of {H}$_2$ by Carrier-Envelope Phase Stabilized Laser Pulses},
  author={Kremer, Manuel and Fischer, Bettina and Feuerstein, Bernold and De Jesus, Vitor L B and Sharma, Vandana and Hofrichter, Christian and Rudenko, Artem and Thumm, Uwe and Schr{\"o}ter, Claus Dieter and Moshammer, Robert and others},
  journal={Phys. Rev. Lett.},
  volume={103},
  number={21},
  pages={213003},
  year={2009},
  publisher={APS}
}

@article{bian2011subcycle,
  title={Subcycle interference dynamics of time-resolved photoelectron holography with midinfrared laser pulses},
  author={Bian, Xue-Bin and Huismans, Y and Smirnova, O and Yuan, Kai-Jun and Vrakking, M J J and Bandrauk, Andr{\'e} D},
  journal={Phys. Rev. A},
  volume={84},
  number={4},
  pages={043420},
  year={2011},
  publisher={APS}
}

@article{meckel2014signatures,
  title={Signatures of the continuum electron phase in molecular strong-field photoelectron holography},
  author={Meckel, M and Staudte, A and Patchkovskii, S and Villeneuve, D M and Corkum, P B and D{\"o}rner, R and Spanner, M},
  journal={Nat. Phys.},
  volume={10},
  number={8},
  pages={594--600},
  year={2014},
  publisher={Nature Publishing Group UK London}
}

@article{ullrich2003recoil,
  title={Recoil-ion and electron momentum spectroscopy: reaction-microscopes},
  author={Ullrich, Joachim and Moshammer, Robert and Dorn, Alexander and D{\"o}rner, Reinhard and Schmidt, L Ph H and Schmidt-B{\"o}cking, H},
  journal={Rep. Prog. Phys.},
  volume={66},
  number={9},
  pages={1463},
  year={2003},
  publisher={IOP Publishing}
}

@article{wu2013electron,
    title={Electron-nuclear energy sharing in above-threshold multiphoton dissociative ionization of {H}$_2$},
    author={Wu, J and Kunitski, M and Pitzer, M and Trinter, F and Schmidt, L Ph H and Jahnke, T and Magrakvelidze, Maia and Madsen, Christian B and Madsen, L B and Thumm, U and others},
    journal={Phys. Rev. Lett.},
    volume={111},
    number={2},
    pages={023002},
    year={2013},
    publisher={APS}
}

@article{Spanner2004,
   author = {Michael Spanner and Olga Smirnova and Paul B Corkum and Misha Yu Ivanov},
   doi = {10.1088/0953-4075/37/12/l02},
   issn = {0953-4075},
   issue = {12},
   journal = {J. Phys. B},
   pages = {L243-L250},
   publisher = {IOP Publishing},
   title = {Reading diffraction images in strong field ionization of diatomic molecules},
   volume = {37},
   url = {http://dx.doi.org/10.1088/0953-4075/37/12/L02},
   year = {2004}
}

@article{Duerr1998,
   author = {S Dürr and T Nonn and G Rempe},
   doi = {10.1038/25653},
   issn = {1476-4687},
   issue = {6697},
   journal = {Nature},
   pages = {33-37},
   title = {Origin of quantum-mechanical complementarity probed by a ‘which-way’ experiment in an atom interferometer},
   volume = {395},
   url = {https://doi.org/10.1038/25653},
   year = {1998}
}

@article{Bienfait2020,
  title = {Quantum Erasure Using Entangled Surface Acoustic Phonons},
  author = {Bienfait, A. and Zhong, Y. P. and Chang, H.-S. and Chou, M.-H. and Conner, C. R. and Dumur, \'E. and Grebel, J. and Peairs, G. A. and Povey, R. G. and Satzinger, K. J. and Cleland, A. N.},
  journal = {Phys. Rev. X},
  volume = {10},
  issue = {2},
  pages = {021055},
  numpages = {8},
  year = {2020},
  month = {Jun},
  publisher = {American Physical Society},
  doi = {10.1103/PhysRevX.10.021055},
  url = {https://link.aps.org/doi/10.1103/PhysRevX.10.021055}
}

@article{Kwiat1992Observation,
  title = {Observation of a ``quantum eraser'': A revival of coherence in a two-photon interference experiment},
  author = {Kwiat, Paul G. and Steinberg, Aephraim M. and Chiao, Raymond Y.},
  journal = {Phys. Rev. A},
  volume = {45},
  issue = {11},
  pages = {7729--7739},
  numpages = {0},
  year = {1992},
  month = {Jun},
  publisher = {American Physical Society},
  doi = {10.1103/PhysRevA.45.7729},
  url = {https://link.aps.org/doi/10.1103/PhysRevA.45.7729}
}

@article{Weisz2014,
   author = {E Weisz and H K Choi and I Sivan and M Heiblum and Y Gefen and D Mahalu and V Umansky},
   issue = {6190},
   journal = {Science},
   month = {6},
   pages = {1363-1366},
   publisher = {American Association for the Advancement of Science},
   title = {An electronic quantum eraser},
   volume = {344},
   url = {https://doi.org/10.1126/science.1248459},
   year = {2014}
}

@article{ruberti2019onset,
  title={Onset of ionic coherence and ultrafast charge dynamics in attosecond molecular ionisation},
  author={Ruberti, M},
  journal={Phys. Chem. Chem. Phys.},
  volume={21},
  number={32},
  pages={17584--17604},
  year={2019},
  publisher={The Royal Society of Chemistry}
}

@article{ruberti2022quantum,
  title={Quantum coherence in molecular photoionization},
  author={Ruberti, Marco and Patchkovskii, Serguei and Averbukh, Vitali},
  journal={Phys. Chem. Chem. Phys.},
  volume={24},
  number={33},
  pages={19673--19686},
  year={2022},
  publisher={The Royal Society of Chemistry}
}

@article{koll2026entanglement,
  title={Entanglement and electronic coherence in attosecond molecular photoionization},
  author={Koll, L-M and Su{\~n}er-Rubio, AJ and Witting, T and Bello, RY and Palacios, A and Mart{\'\i}n, Fernando and Vrakking, MJJ},
  journal={Nature},
  volume={652},
  number={8108},
  pages={82--88},
  year={2026},
  publisher={Nature Publishing Group UK London}
}

@article{spathe2025laser,
  title={Laser-driven high-flux source of coherent quasi-monochromatic extreme ultraviolet radiation for coincidence spectroscopy},
  author={Sp{\"a}the, Julian and Hell, Sebastian and W{\"u}nsche, Martin and Klas, Robert and Rothhardt, Jan and Limpert, Jens and Siefke, Thomas and Paulus, Gerhard G and K{\"u}bel, Matthias},
  journal={Review of scientific instruments},
  volume={96},
  number={12},
  year={2025},
  publisher={AIP Publishing}
}

@article{becker2002above,
author = {W Becker and F Grasbon and R Kopold and D B Milo{\v{s}}evi{\'c} and G G  Paulus and H Walther},
title = {Above-Threshold Ionization: From Classical Features to Quantum Effects},
journal = {Adv. At. Mol. Opt. Phys.},
publisher = {Academic Press},
volume = {48},
pages = {35-98},
year = {2002},
issn = {1049-250X},
doi = {https://doi.org/10.1016/S1049-250X(02)80006-4},
url = {https://www.sciencedirect.com/science/article/pii/S1049250X02800064}
}

@article{Corkum1993plasma,
  title = {Plasma perspective on strong field multiphoton ionization},
  author = {Corkum, P. B.},
  journal = {Phys. Rev. Lett.},
  volume = {71},
  issue = {13},
  pages = {1994--1997},
  numpages = {0},
  year = {1993},
  month = {Sep},
  publisher = {American Physical Society},
  doi = {10.1103/PhysRevLett.71.1994},
  url = {https://link.aps.org/doi/10.1103/PhysRevLett.71.1994}
}

@article{bucksbaum1990softening,
  title={Softening of the {H}$_2^+$ molecular bond in intense laser fields},
  author={Bucksbaum, Phil H and Zavriyev, Anton and Muller, Harm G and Schumacher, Douglass W},
  journal={Phys. Rev. Lett.},
  volume={64},
  number={16},
  pages={1883},
  year={1990},
  publisher={APS}
}

@article{huismans2011time,
  title={Time-resolved holography with photoelectrons},
  author={Huismans, Ymkje and Rouz{\'e}e, Arnaud and Gijsbertsen, Arjan and Jungmann, J H and Smolkowska, A S and Logman, P S W M and Lepine, Franck and Cauchy, C{\'e}cile and Zamith, S{\'e}bastien and Marchenko, Tatiana and others},
  journal={Science},
  volume={331},
  number={6013},
  pages={61--64},
  year={2011},
  publisher={American Association for the Advancement of Science}
}

@article{lu2018high,
  title={High-order above-threshold dissociation of molecules},
  author={Lu, Peifen and Wang, Junping and Li, Hui and Lin, Kang and Gong, Xiaochun and Song, Qiying and Ji, Qinying and Zhang, Wenbin and Ma, Junyang and Li, Hanxiao and others},
  journal={Proc. Natl. Acad. Sci. U.S.A.},
  volume={115},
  number={9},
  pages={2049--2053},
  year={2018},
  publisher={National Acad Sciences}
}

@article{He2020,
  title = {Laser-wavelength and intensity dependence of electron-nuclear energy sharing in dissociative ionization of {H}$_{2}$},
  author = {He, Chaoxiong and Liang, Hao and Liu, Ming-Ming and Peng, Liang-You and Liu, Yunquan},
  journal = {Phys. Rev. A},
  volume = {101},
  issue = {5},
  pages = {053403},
  numpages = {7},
  year = {2020},
  month = {May},
  publisher = {American Physical Society},
  doi = {10.1103/PhysRevA.101.053403},
  url = {https://link.aps.org/doi/10.1103/PhysRevA.101.053403}
}

@article{Laurell2025,
   author = {Hugo Laurell and Sizuo Luo and Robin Weissenbilder and Mattias Ammitzböll and Shahnawaz Ahmed and Hugo Söderberg and C Leon. M Petersson and Vénus Poulain and Chen Guo and Christoph Dittel and Daniel Finkelstein-Shapiro and Richard J Squibb and Raimund Feifel and Mathieu Gisselbrecht and Cord L Arnold and Andreas Buchleitner and Eva Lindroth and Anton Frisk Kockum and Anne L’Huillier and David Busto},
   doi = {10.1038/s41566-024-01607-8},
   issn = {1749-4893},
   issue = {4},
   journal = {Nat. Photon.},
   pages = {352-357},
   title = {Measuring the quantum state of photoelectrons},
   volume = {19},
   url = {https://doi.org/10.1038/s41566-024-01607-8},
   year = {2025}
}

@article{Walborn2002,
   author = {S P Walborn and M O Terra Cunha and S Pádua and C H Monken},
   doi = {10.1103/PhysRevA.65.033818},
   issue = {3},
   journal = {Phys. Rev. A},
   month = {2},
   pages = {33818},
   publisher = {American Physical Society},
   title = {Double-slit quantum eraser},
   volume = {65},
   url = {https://link.aps.org/doi/10.1103/PhysRevA.65.033818},
   year = {2002}
}

@article{Scully1991,
   author = {Marian O Scully and Berthold-Georg Englert and Herbert Walther},
   doi = {10.1038/351111a0},
   issn = {1476-4687},
   issue = {6322},
   journal = {Nature},
   pages = {111-116},
   title = {Quantum optical tests of complementarity},
   volume = {351},
   url = {https://doi.org/10.1038/351111a0},
   year = {1991}
}

    \begin{appendix}

\section{Dissociation pathways}
    This section describes the interfering dissociative ionization pathways that lead to the formation of the entangled electron-ion state.    
    
    \begin{figure}
        \centering
        \includegraphics[width=\linewidth]{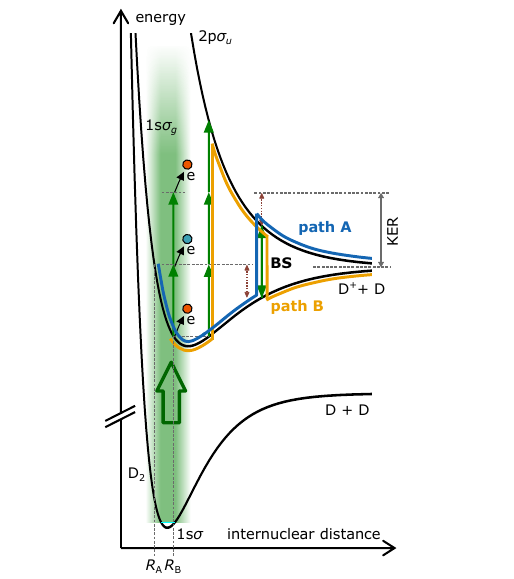}
        \caption{\small\justifying Schematic showing the relevant molecular potential-energy curves (black lines) and the interfering paths A and B, which result in entanglement between photoelectron and ion. The Franck-Condon region is indicated by the green shaded area.}
        \label{fig:diss_pathways}
    \end{figure}

    Figure \ref{fig:diss_pathways} shows the potential-energy curves of $\mathrm{D}_2^+$, and the interfering pathways leading to entanglement between photoelectron and remaining molecular ion. Paths A and B interfere because they lead to identical photoelectron energy and nuclear KER. The existence of interfering paths is possible because the Franck-Condon region in the ${\rm 1s}\sigma_g$ state of $\mathrm{D}_2^+$ spans across a potential-energy range that exceeds one photon energy (\SI{2.4}{eV}). For path B, the transition from the neutral to the cationic ground state (i.e.~ionization) occurs at an internuclear distance $R_\mathrm{B}$ close to the potential minimum of ${\rm 1s}\sigma_g$. For path A, ionization occurs at a smaller internuclear distance $R_\mathrm{A}$ where the potential-energy is precisely one photon energy higher than for path B, i.e.~$V(R_A) = V(R_B) + \hbar \omega$. 
    
    Following ionization, the nuclear wavepacket propagates on the ${\rm 1s}\sigma_g$ state. On path A, the molecular ion is excited to the dissociative ${\rm 2p}\sigma_u$ state by the absorption of one photon. Thus, the ion parity for path A is odd. For path B, the ion dissociates via net-two-photon absorption \cite{giusti1990above}, which involves a three-photon excitation to the ${\rm 2p}\sigma_u$ state and subsequent de-excitation to the ${\rm 1s}\sigma_g$ state. Thus, the ion parity for path B is even. Since the potential energies at the initial internuclear distances ($R_\mathrm{A}$, $R_\mathrm{B}$) differ by one photon energy, the photoelectron parities for the two paths are different. This leads to the entangled electron-ion state described by Eq.~(\ref{eq:psi_ug_basis}).

\section{Transformation from parity basis to left/right basis \label{sec:trafo_basis}}

This section describes how the overlap of even ($\ket{g}$) and odd ($\ket{u}$) parity contributions causes a localization of the wave function on the left or right side in space, leading to a predominant directional emission of both, the ion and the photoelectron.

For the $\mathrm{D}_2^+$ ion, the relevant parity states are $\ket{g}_\mathrm{D_2^+} = \ket{{\rm 1s}\sigma_g}$ and $\ket{u}_\mathrm{D_2^+} = \ket{{\rm 2p}\sigma_u}$. For the photoelectron, produced by multiphoton ionization, $\ket{g}_\mathrm{e}$ and $\ket{u}_\mathrm{e}$ consist of sums of even and odd spherical harmonics, respectively. 

To analyze the spatial localization of photoelectron and ion, we transform to a left/right basis: $\ket{l} = (\ket{g} - \ket{u})/\sqrt{2}$ and $\ket{r} = (\ket{g} + \ket{u})/\sqrt{2}$. Substituting these into Eq.~(\ref{eq:psi_ug_basis}) yields the entangled ion-photoelectron state in the $r/l$ basis 
\begin{equation}
    \begin{aligned}
    \ket{\psi_{\mathrm{A+B}}} =
    [ (\alpha + \beta e^{i\phi})(&\ket{r, r} \pm \ket{l,l}) \\
    + (\beta e^{i\phi} - \alpha)(&\ket{r, l} \pm \ket{l,r}) ]  / 2 \; ,
    \end{aligned}
    \label{eq:psi_rl_basis}
\end{equation}
where an even (odd) number of absorbed photons results in the sign $+$ ($-$) and the two-particle state is denoted as $\ket{\mathrm{photoelectron},\mathrm{D}_2^+}$. The localization of the bound electron in D$_2^+$ on one side results in a consequent localization of the positive charge on the other side. Consequently, one can rewrite Eq.~(\ref{eq:psi_rl_basis}) using $\ket{r}_\text{D$_2^+$} = \ket{L}_{\mathrm{D}^+}$ and $\ket{l}_\text{D$_2^+$} = \ket{R}_{\mathrm{D}^+}$, which results in Eq.~(\ref{eq:psi_rL_basis}). 
Crucially, this spatial localization on the left or right side dictates the emission direction. During dissociation, the localization of the bound electron on one side means that the $\mathrm{D}^+$ ion moves into the opposite direction. Similarly, for the photoelectron, the interference of even and odd parity contributions, underlying spatial localization, typically implies a preferred emission direction \cite{yin1992asymmetric, lafosse2003molecular, martin2007single}. 

\section{Entanglement evidence \label{sec:entanglement_evidence}}

    This section describes in detail how the non-zero emission correlation parameter $C$ provides evidence for entanglement between photoelectron and dissociating molecular ion. Subsection 1 describes why a non-zero emission correlation is evidence of coherence in a  two-particle state and thus for entanglement, provided the total parity is either even or odd. In subsection 2, it is shown that $C$ is proportional to the fidelity (\textit{i.e.} overlap) of the two-particle state with an ideal Bell state, which makes the measured oscillation in $C$ with KER an evidence of entanglement. Subsection 3 discusses the implicit assumptions and the limitations of the presented entanglement evidence. Finally, subsection 4 discusses the appearance of non-zero emission direction correlation in the bond-softening (BS) region of the KER distribution.  
    
    \subsection{Correlated electron-ion emission as evidence of two-particle coherence}
    To show that the electron-ion emission direction correlation indeed provides evidence of entanglement, we must consider all possible states and calculate their expected emission direction correlation. We define the directional correlation $C$ as:
    \begin{equation}
        C = \frac{(p_{r,R} + p_{l,L}) - (p_{r,L} + p_{l,R})}{(p_{r,R} + p_{l,L}) + (p_{r,L} + p_{l,R})} \; ,
        \label{eq:asym_theory}
    \end{equation}
    where $p_{r,R}$, for instance, is the probability of both $\mathrm{D^+}$ ion and photoelectron being emitted to the right, as obtained by projecting onto the state $\ket{r,R}$. Here, perfect spatial left/right localization is assumed in order to simplify the derivation. In case of the ion, one can safely assume perfect localization once the dissociation is complete. This assumption does not hold for the photoelectron. Hence, there is a non-zero probability for the photoelectron to be emitted to right (left) if it is in state $\ket{l}$ ($\ket{r}$). Consequently, the equations for $C$, derived in this section, provide an upper limit and the magnitude of the actual observable correlation $C$ is smaller.
    
    While a general, pure separable two-qubit state such as  
    \begin{equation}
        \ket{\psi_\mathrm{sep}} = \underbrace{(\zeta_g\ket{g} + \zeta_u\ket{u})}_\text{photoelectron} \otimes \underbrace{(\eta_g\ket{g} + \eta_u\ket{u})}_{\mathrm{D}_2^+} 
        \label{eq:psi_sep}
    \end{equation}
    can mathematically produce a non-zero emission direction correlation 
    \begin{equation}
        C = -2[\mathrm{Re}(\zeta_g^* \zeta_u \eta_g^* \eta_u) + \mathrm{Re}(\zeta_g^* \zeta_u \eta_g \eta_u^*)] \, ,
        \label{eq:asym_separable_state}
    \end{equation}
    this requires a superposition of different total parities. In our experiment, however, the total parity is fixed for a given kinetic energy release (KER) and photoelectron energy $E_\mathrm{e}$. This holds with the exception to the vicinity of the BS peak, as discussed in section \ref{section:bs_correlation}. As seen in Fig.~\ref{fig:asymmetry}(a), diagonal lines in the joint energy spectrum signify a defined total absorbed energy, fixing the number of absorbed photons and thus the overall parity. 
    
    Restricting our focus to the KER region between \SI{2}{\eV} and \SI{3}{\eV}, the two-particle state has a well-defined parity. The density matrix for an even parity state ($\rho_{+1}$), for example, is spanned exclusively by $\ket{g,g}$ and $\ket{u,u}$ (analogously, $\rho_{-1}$ is spanned by $\ket{g,u}$ and $\ket{u,g}$):
    \begin{equation}
        \rho_{+1} = 
        \begin{pmatrix}
            p_{gg} & 0 & 0 & c_{+1} \\
            0 & 0 & 0 & 0 \\
            0 & 0 & 0 & 0 \\
            c^*_{+1} & 0 & 0 & p_{uu}
        \end{pmatrix} \, , \;
        \rho_{-1} = 
        \begin{pmatrix}
            0 & 0 & 0 & 0 \\
            0 & p_{gu} & c_{-1} & 0 \\
            0 & c_{-1}^* & p_{ug} & 0 \\
            0 & 0 & 0 & 0
        \end{pmatrix} \; .
        \label{eq:density_parity_conserved_state}
    \end{equation}
    Here, $p_{gg}$, for instance, is the probability to detect the state $\ket{g,g}$ and $c_{+1}$ is the coherence between $\ket{g,g}$ and $\ket{u,u}$ in the state described by $\rho_{+1}$. For an odd-parity pure state $\ket{\psi_\text{odd}}=\alpha \ket{g,u} + \beta e^{i\phi}\ket{u,g}$ with real positive $\alpha$ and $\beta$, for example, the probabilities and coherence in the respective density matrix are given by $p_{gu}=\alpha^2$, $p_{ug}=\beta^2$ and $c_{-1}=\alpha\beta e^{-i\phi}$. 
    
    Evaluating $C$ for these fixed-parity density matrices in the left/right basis yields:
    \begin{equation}
        C(\rho_{\pm 1}) = -2 \mathrm{Re}(c_{\pm 1}) \; ,
        \label{eq:asym_theory_parity_conserved}
    \end{equation}
    where $c_{\pm 1}$ is the coherence between the allowed states for a fixed total parity (\textit{e.g.}, between $\ket{g,g}$ and $\ket{u,u}$). For an incoherent statistical mixture $c_{\pm 1}=0$ and thus the correlation $C$ vanishes. Conversely, for a pure, entangled state such as $\ket{\psi_\mathrm{A+B}} = \alpha \ket{g,g} + \beta e^{i\phi}\ket{u,u}$, the correlation becomes:
    \begin{equation}
        C(\ket{\psi_\mathrm{A+B}}) = -2\alpha \beta \cos\phi \; .
        \label{eq:asym_theory_entangl}
    \end{equation}
    
    The relative phase $\phi$ varies with KER due to the differing shapes of the ${\rm 1s}\sigma_g$ and ${\rm 2p}\sigma_u$ potential-energy curves (paths A and B). While a variation in the population amplitudes [$\alpha$ or $\beta$, as hinted in Fig.~\ref{fig:intro}(a)] could induce fluctuations in $C$, these amplitudes are strictly non-negative and would only produce oscillations around a non-zero offset. Therefore, the oscillation of $C$ around zero observed in Fig.~\ref{fig:asymmetry}(b) must originate from the changing relative phase $\phi$. This provides direct, measurable evidence of two-body coherence and, consequently, entanglement between the photoelectron and the dissociating ion.

    \subsection{Fidelity with maximally entangled Bell states as an entanglement evidence}

    In order to obtain a quantitative measure of entanglement, one has to calculate the fidelity $F$ (overlap) with the Bell states, which are a normalized basis set for a maximally entangled two-particle system. The Bell states for the parity qubit states $\ket{g}$ and $\ket{u}$ are defined as:
\begin{equation}
\begin{aligned}
    \ket{\Phi^\pm} &= \tfrac{1}{\sqrt{2}}(\ket{g}\otimes\ket{g} \pm \ket{u}\otimes\ket{u}) \\
    \ket{\Psi^\pm} &= \tfrac{1}{\sqrt{2}}(\ket{g}\otimes\ket{u} \pm \ket{u}\otimes\ket{g}) \; .
\end{aligned}
    \label{eq:bell_states}
\end{equation}

For any pure or mixed separable (non-entangled) state, the Cauchy-Schwarz inequality restricts the fidelity with any Bell state to a classical upper limit of $F_\mathrm{sep} \leq 1/2$. 

While a fidelity $F > 1/2$ is a sufficient condition for entanglement, not all mathematically possible entangled states exceed this threshold. However, parity conservation in our experiment restricts the accessible state space, ensuring that our observable directly measures this fidelity limit. By detecting the KER and photoelectron energy $E_\mathrm{e}$, the total number of absorbed photons is known, yielding a well-defined even ($+1$) or odd ($-1$) total parity for the system. 

For the resulting fixed-parity density matrices [$\rho_{+1}$ or $\rho_{-1}$, as defined in Eq.~(\ref{eq:density_parity_conserved_state})], the fidelity with the corresponding closest Bell state simplifies significantly:
\begin{equation}
\begin{aligned}
    F_{+1} &= \bra{\Phi^\pm}\rho_{+1}\ket{\Phi^\pm} = \tfrac{1}{2} \pm \mathrm{Re}(c_{+1}) \\
    F_{-1} &= \bra{\Psi^\pm}\rho_{-1}\ket{\Psi^\pm} = \tfrac{1}{2} \pm \mathrm{Re}(c_{-1}) \; .
\end{aligned}
\end{equation}
Here, we utilized the normalization of the state populations (\textit{e.g.}, $p_{gg} + p_{uu} = 1$). 

Therefore, under the condition of parity conservation, any two-particle coherence ($c_{+1}$ or $c_{-1}$) with a non-zero real part drives the fidelity above the classical limit of $1/2$. As established in Eq.~(\ref{eq:asym_theory_parity_conserved}), our experiment directly accesses the real part of this coherence via the emission direction correlation $C$. Consequently, measuring a non-zero correlation ($C \neq 0$) serves as direct, quantitative evidence of entanglement, demonstrating that the fidelity with a maximally entangled Bell state exceeds the classical threshold.

    \subsection{Validity of entanglement evidence via emission direction correlation}

While the fact that a non-entangled, incoherent state exhibits no emission direction correlation ($C=0$, assuming parity conservation) might suggest that any decrease in $|C|$ reflects a loss of coherence and entanglement, several inherent physical factors actually limit the maximum observable correlation. 

Specifically, the expected correlation is bounded by the following constraints:
\begin{itemize}
    \item \textbf{Varying state amplitudes:} The magnitudes of $\alpha$ and $\beta$ change with KER [as seen in Fig.~\ref{fig:intro}(a)]. A perfect correlation ($C=1$) would strictly require an equal superposition ($\alpha=\beta=1/\sqrt{2}$).
    \item \textbf{Incomplete spatial localization:} Overlapping opposite-parity contributions cause predominant, rather than exclusive, emission in one direction. Thus, perfect parity correlation does not translate to perfect emission direction correlation, a limitation confirmed by our TDSE calculations. 
    \item \textbf{Strong-field effects:} Additional dissociation and ionization pathways or incoherence introduced by the intense laser field can further dampen the observable correlation.
\end{itemize}

Despite these limiting factors, the measured correlation amplitude [Fig.~\ref{fig:asymmetry}(b)] agrees qualitatively well with the results of the TDSE calculations. The reduced magnitude of the measured correlation compared with the TDSE calculation may result from effects not included in the model:
First, the TDSE calculations were performed for fixed intensity values and neglect the averaging over the focal intensity distribution in the experiment. As shown in Fig.~\ref{fig:asym_all_intensities}, the phase of the KER-dependent oscillations in $C$ depends sensitively on the laser intensity. Averaging over the experimental intensity distribution in the focal volume therefore superposes oscillations with different phases, leading to a significantly reduced modulation amplitude of $C$. Second, the TDSE restricts the treatment of the bound electron in D$_2^+$ to the two lowest electronic states, ${\rm 1s}\sigma_g$ and ${\rm 2p}\sigma_u$. Additional dissociation pathways may therefore reduce the experimentally observed correlation. Third, the reduced dimensionality of the TDSE can modify the PMD \cite{brennecke2020gouy}, which may alter the relationship between spatial localization and emission direction and therefore potentially contributes to the larger C observed in the TDSE.

Nonetheless, the oscillation in the correlation as a function of KER and the intensity dependent phase-shift of these oscillations is qualitatively well reproduced by the TDSE (compare Fig.~\ref{fig:asym_all_intensities}). This agreement underscores the robustness of our entanglement evidence and indicates that experimental incoherence plays only a minor role.

Furthermore, we must rule out technical artifacts, such as asymmetric detector efficiencies or laser-induced biases. We exclude these by demonstrating that $C$ oscillates symmetrically around zero as a function of the relative phase $\phi$ between dissociation paths A and B. Although tuning $\phi$ directly at a fixed KER is experimentally prohibitive, the differing curvatures of the ${\rm 1s}\sigma_g$ and ${\rm 2p}\sigma_u$ states (Fig.~\ref{fig:diss_pathways}) cause the accumulated phase $\phi$ to naturally vary with KER. Thus, filtering by KER effectively tunes $\phi$.

This method of scanning $\phi$ relies on two core assumptions:
\begin{enumerate}
    \item Detecting the KER and photoelectron energy determines the number of absorbed photons, and thus fixes the total parity of the system.
    \item The states across different KERs are prepared identically (maintaining the same parity purity) up to the phase shift $\phi$. This is directly analogous to the assumption of identical state preparation required when measuring sequential copies in a standard Bell test.
\end{enumerate}

Given these conditions, the observed symmetric oscillation of the emission direction correlation demonstrates two-particle coherence. Consequently, the fidelity of the detected electron-ion state with a maximally entangled Bell state exceeds the classical limit, providing conclusive evidence of entanglement.

    \subsection{Emission direction correlation in the bond-softening regime \label{section:bs_correlation}}
    
    
    As shown in Fig.~\ref{fig:intro}(a), the contribution of path A (${\rm 2p}\sigma_u$ state) dominates over path B below \SI{1.8}{\eV} KER, thus $\alpha \gg \beta$. Consequently, Eq.~(\ref{eq:asym_theory_entangl}) predicts a smaller emission direction correlation amplitude than in regions where $\alpha \approx \beta$. The TDSE results agrees with this expectation, showing little correlation below \SI{1.8}{\eV} [Fig.~\ref{fig:asymmetry}(b)]. 

    Experimentally, however, the emission correlation $C$ below \SI{1.8}{\eV} is larger than the TDSE predicts, suggesting that the two-state Born-Oppenheimer approximation does not fully describe the dynamics of the molecular ion in the low-KER region. This enhanced correlation could arise from two sources. First, non-adiabatic dynamics might increase the ${\rm 1s}\sigma_g$ state's contribution. However, the measured $C$ lacks the clear KER-dependent oscillations seen at higher KERs, suggesting that the non-zero $C$ has a different physical origin. Second, the system might form a separable state below \SI{1.8}{\eV}, which, according to Eq.~(\ref{eq:asym_separable_state}), can also yield a non-zero emission direction correlation.

    Forming a separable state [Eq.~(\ref{eq:psi_sep})] requires even and odd parity contributions to overlap at identical KER and photoelectron energies. Indeed, previous work using few-cycle pulses shows that the ion emission direction in the BS regime correlates with the carrier-envelope phase  \cite{kremer2009electron}, implying parity mixing in the D$_2^+$ ion. Above the BS regime ($\mathrm{KER}>\SI{1.8}{\eV}$), photoelectron and KER are correlated, as indicated by the diagonal lines in Fig.~\ref{fig:asymmetry}(a), and thus the sum $\mathrm{KER}+E_\mathrm{e}$ encodes the number of absorbed photons and thus the photoelectron's parity. As this energy correlation is lost in the BS regime, parity cannot be deduced from measuring KER and $E_\mathrm{e}$, meaning that mixed photoelectron parity is possible.
    
    While parity mixing in the BS regime yields a non-zero emission direction correlation, one might expect it to simultaneously wash out photoelectron holography. However, these two observables have different sensitivities: emission correlation scales linearly with the product $\alpha\beta$, whereas the loss of holography requires the amplitudes to be nearly balanced ($|\alpha|^2 \approx |\beta|^2$). Therefore, holography is much more sensitive to amplitude imbalances between pathways A and B. This explains why Fig.~\ref{fig:asymmetry}(b) displays pronounced correlation oscillations up to \SI{3}{\eV} KER, even though above \SI{2.3}{\eV}, $\beta$ exceeds $\alpha$ by an order of magnitude [Fig.~\ref{fig:intro}(a)] and holography is restored. As ${\rm 2p}\sigma_u$ is expected to dominate in the BS regime, the clear visibility of the holography in the PMD does not contradict the parity mixing in the BS regime.

\section{Time-domain interpretation of the existence of which-way information}

    \begin{figure}
        \centering
        \includegraphics[width=\linewidth]{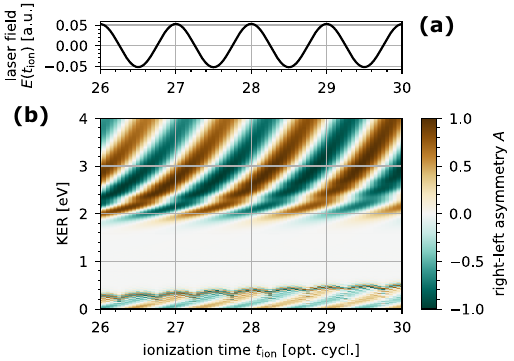}
        \caption{\small \justifying 
            (a) Laser pulse electric field $E(t_0)$ in atomic units at the respective ionization time $t_0$ for the TDSE results of the lower subfigure. 
            (b) Normalized right-left ion emission direction asymmetry $A$ [Eq.~(\ref{eq:left-right-asym})] in the lab frame, as obtained from a 1D ion-only TDSE calculation of the D$_2^+$ dissociation within the two-level BO approximation for different ionization times $t_\mathrm{ion}$. Each column corresponds to an individual TDSE simulation, where the nuclei is initialized in a Franck-Condon state for this particular moment of ionization within the laser pulse. A small $t_\mathrm{ion}$ corresponds to early ionization in the pulse, a large $t_\mathrm{ion}$ to later ionization. The laser intensity is \SI{9.74e13}{W/cm^2} for this calculation.}
        \label{fig:1d_TDSE_asym_ion}
    \end{figure}

    In this section, we show that the D$^+$ ion emission direction provides which-way information about the sign of the tunnel exit position.

    The success of the simple interference model using two Gaussian wavepackets to explain the loss of holography implies that the two tunnel exit points act as the slits in our molecular quantum eraser. Therefore, knowing the sign of the laser's electric field at the moment of ionization provides which-way information about the photoelectron's path.

    To determine if the dissociating ion retains information about the ionization time, we perform a 1D, ion-only TDSE calculation. We model the dissociation of D$_2^+$ within the two-level Born-Oppenheimer (BO) approximation. The time propagation begins at the moment of ionization ($t_\mathrm{ion}$), which occurs near the field maximum at 28 optical cycles (o.c.) within a 56 o.c. (\SI{35}{fs} FWHM) pulse. The nuclei start in a Franck-Condon state on the $\mathrm{1s}\sigma_g$ potential curve and are propagated to the end of the pulse. We then calculate the right-left ion emission asymmetry
    \begin{equation}
    A = \frac{\left|\braket{\psi_{\mathrm{D}_2^+}}{R}\right|^2 - \left|\braket{\psi_{\mathrm{D}_2^+}}{L}\right|^2}{\left|\braket{\psi_{\mathrm{D}_2^+}}{R}\right|^2 + \left|\braket{\psi_{\mathrm{D}_2^+}}{L}\right|^2}
        \label{eq:left-right-asym}
    \end{equation}
    of the dissociated nuclei in the lab frame as a function of the kinetic energy release (KER). The TDSE calculations are repeated for different values of $t_\mathrm{ion}$. The calculated right-left asymmetry $A$ is shown in Fig.~\ref{fig:1d_TDSE_asym_ion} as a function of KER and $t_\mathrm{ion}$.


    As shown in Fig.~\ref{fig:1d_TDSE_asym_ion}, a strong ion emission asymmetry emerges above \SI{1.8}{\eV} KER, oscillating as a function of $t_\mathrm{ion}$ at the frequency of the laser. Because $t_\mathrm{ion}$ dictates the sign of the laser electric field, the ion's emission direction clearly correlates with this sign; and hence with the sign of the tunnel exit position. Furthermore, since the emission direction is tied to the relative phase $\phi$ between dissociation paths A and B, $\phi$ also correlates with the laser field sign. Consequently, for $\mathrm{KER} > \SI{1.8}{\eV}$, the dissociating ion encodes which-way information regarding the left or right tunnel exit position of the photoelectron.

\section{KER spectra for different laser intensities and pulse durations \label{sec:KER_spectra_diff_int_diff_pulse_dur}}

    \begin{figure}
        \centering
        \includegraphics[width=\linewidth]{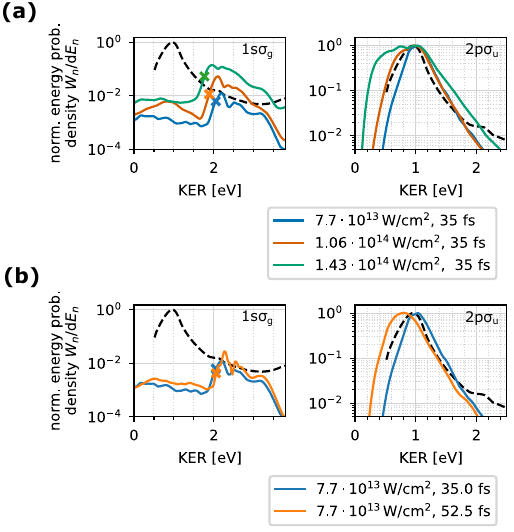}
        \caption{\small \justifying 
            Kinetic energy release (KER) spectra of the ${\rm 1s}\sigma_g$ (left) and the ${\rm 2p}\sigma_u$ state (right) for different TDSE calculations. The KER spectra are normalized to the maximum of the KER spectrum of the ${\rm 2p}\sigma_u$ state for each TDSE calculation to be comparable to the experimental KER spectrum. Without normalization, the overal energy prob. density increases significantly with laser intensity and marginally with laser pulse duration. For comparison, the normalized KER spectrum of the experiment presented in Fig.~\ref{fig:PMD} (\SI{9e13}{W/cm^2} laser intensity and \SI{35}{fs} pulse duration) is shown as a dashed black line. As indicated in the figure legend, calculated data for (a) different laser intensities and (b) different pulse durations (FWHM) are shown.}
        \label{fig:KER_all_intensities}
    \end{figure}
    
    As mentioned in the methods section, we performed multiple full 3D TDSE calculations at high resolution to determine which laser parameters (\textit{i.e.} peak intensity and pulse duration) have to be chosen for the TDSE model to best reproduce the experimental data. The normalized KER spectra obtained from these calculations are plotted in Fig.~\ref{fig:KER_all_intensities} separately for dissociation on the ${\rm 1s}\sigma_g$ state (left panel) and for dissociation on the ${\rm 2p}\sigma_u$ state (right panel). Subfigure (a) shows calculations for three different laser intensities at the same pulse duration, whereas subfigure (b) shows two calculations with different laser pulse durations but for the same intensity. 
    
    Figure~\ref{fig:KER_all_intensities} shows that increasing the laser intensity enhances the relative contribution of dissociation on the ${\rm 1s}\sigma_g$ state and shifts it toward lower KER values. At the same time, the BS peak associated with dissociation on the ${\rm 2p}\sigma_u$ state broadens asymmetrically towards ö, while its maximum position remains largely unchanged. In contrast, increasing the pulse duration shifts the BS peak toward lower KER values, whereas the ${\rm 1s}\sigma_g$ contribution moves only slightly toward higher KER and increases only marginally in relative strength.
    
    We selected the calculations for an intensity of \SI{7.7e13}{W/cm^2} and a pulse duration of \SI{52.5}{fs} to be compared to the experiment in the main part of the paper, since these TDSE calculations fit best to the experimental data, especially around a KER of \SI{2}{eV} where the holographic interference is lost in the experimental PMD shown in Fig.~\ref{fig:PMD}. 
    Due to experimental uncertainty in the knowledge of laser intensity and pulse duration and the absence of focal volume averaging in the TDSE, it is expected that the best agreement between TDSE and experiment might be for laser parameters in the TDSE which differ from the experimental ones.

\section{Emission correlation for all measured and calculated laser intensities}

    \begin{figure}
        \centering
        \includegraphics[width=\linewidth]{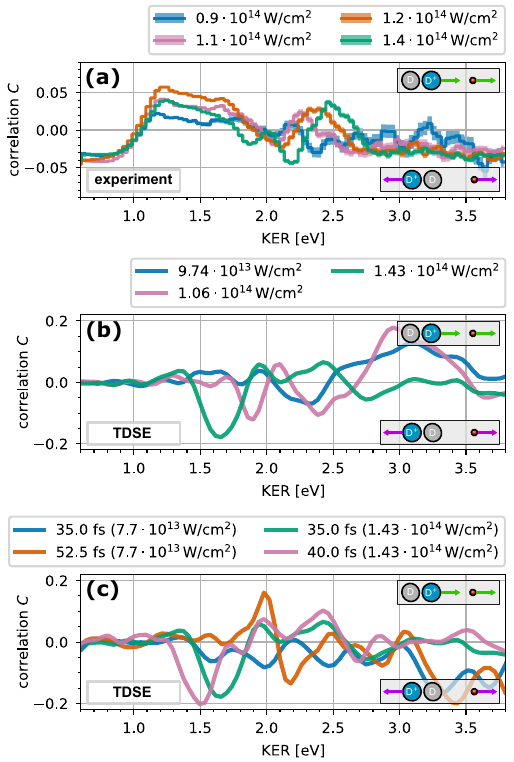}
        \caption{\small \justifying 
            Emission correlation parameter $C$ as a function of KER for (a) experiment and (b,c) TDSE. For the experiment, $C$ is calculated using Eq.~\ref{eq:correlation_exp}. The correlation $C$ is calculated for the TDSE by filtering the wavefunction to correlated and anti-correlated emission directions and integrating the probability density (\textit{i.e.} absolute squared of the wavefunction) over the 2D photoelectron momentum. As indicated in the figure legend, data for (a,b) different laser intensities (but same pulse duration of \SI{35}{fs}) and (c) different pulse durations (FWHM) are shown. A moving average filter (width = three bins $\approx \SI{0.104}{\eV}$) is used to reduce noise in the experimental data. For comparability, the TDSE uses the same moving average filter. The shaded area indicates the propagated standard deviation ($\sigma_C$), assuming Poissonian counting statistics ($\sigma_N=\sqrt{N}$ for $N$ coincidence events). }
        \label{fig:asym_all_intensities}
    \end{figure}
    
    Figure~\ref{fig:asym_all_intensities} expands on the main text by presenting the (a) measured  and (b,c) calculated emission correlation parameter $C$, for different laser intensities and pulse durations. This extended dataset clearly demonstrates two intensity-dependent trends in $C$: an increase in the amplitude of the KER-dependent oscillations, and a distinct phase shift. The larger oscillation amplitude reflects a change in the relative contributions of path A and B at higher intensities, while the phase shift is attributed to the laser-induced Stark shift distorting the potential-energy curves. Moreover, the calculations presented in Fig.~\ref{fig:asym_all_intensities}(c) indicate that the phase shift of the KER-dependent oscillations in $C$ also depends on the laser pulse duration.

    \end{appendix}

\end{document}